\documentclass[a4paper,11pt]{article}

\usepackage[utf8]{inputenc}
\usepackage[T1]{fontenc}
\usepackage[english]{babel}

\usepackage{geometry} 
\usepackage{marginnote}
\usepackage{mathtools}

\usepackage{graphicx}
\usepackage{amsmath}
\usepackage{amsfonts}
\usepackage{amssymb}
\usepackage{pstricks-add}

\usepackage{multicol,caption}
\usepackage{geometry}
\usepackage{caption}
\usepackage{lipsum}
\usepackage{tabularx}
\usepackage{multirow}
\usepackage{subfigure}
\usepackage{ifthen}

\author{
  Adélaïde Raguin$^{1,2,3}$ , Norbert Kern$^{1,*}$, Andrea Parmeggiani$^{1,2,*}$
  \\[0.5em]
  \begin{minipage}{0.9\textwidth}
  $^1$ Laboratoire Charles Coulomb, CNRS, University of Montpellier, Montpellier, France
\\
$^2$ LPHI, CNRS, INSERM, University of Montpellier, Montpellier, France
\\
$^3$ Institute of Quantitative and Theoretical Biology, Department of Biology, Heinrich-Heine University, Düsseldorf 40225, Germany
\\
$*$ authors provided equal contributions.
\\
Corresponding authors: adelaide.raguin@hhu.de, norbert.kern@umontpellier.fr, andrea.parmeggiani@umontpellier.fr
\\
\\© 2020. This manuscript version is made available under the CC-BY-NC-ND 4.0 license http://creativecommons.org/licenses/by-nc-nd/4.0/
\end{minipage}
}

\title{Stochastic modelling of collective motor protein transport through a crossing of microtubules}

\begin{document}
\maketitle

\newenvironment{Figure}
{\par\medskip\noindent\minipage{\linewidth}}
{\endminipage\par\medskip}

\abstract{
  The cytoskeleton in eukaryotic cells plays several crucial roles.  In terms of intracellular transport, motor proteins use the cytoskeletal filaments as a backbone along which they can actively transport biological cargos such as vesicles carrying biochemical reactants. Crossings between such filaments constitute a key element, as they may serve to alter the destination of such payload. Although motor proteins are known to display a rich behaviour at such crossings, the latter have so far only been modelled as simple branching points. Here we explore a model for a crossing between two microtubules which retains the individual tracks consisting of protofilaments, and we construct a schematic representation of the transport paths. We study collective transport exemplified by the Totally Asymmetric Simple Exclusion Process (TASEP), and provide a full analysis of the transport features and the associated phase diagram, by a generic mean-field approach which we confirm through particle-based stochastic simulations. In particular we show that transport through such a compound crossing cannot be approximated from a coarse-grained structure with a simple branching point. Instead, it gives rise to entirely new and counterintuitive features: the fundamental current-density relation for traffic flow is no longer a single-valued function, and it furthermore differs according to whether it is observed upstream or downstream from the crossing. We argue that these novel features may be directly relevant for interpreting experimental measurements. 

}

\begin{multicols}{2}
  
\section{Introduction}

 Understanding how the logistics inside a cell is run constitutes one of the major challenges in biology, with strong and direct implications for medicine  \cite{hirokawa2010,hirokawa2015, courson2015, heissler2017, vera2019} and biotechnologies \cite{andorfer2019, saper2019}. One central issue is that active transport is required in order to efficiently deliver biomaterials over distances which can no longer be covered by Brownian motion on the required timescale. This is achieved by ATP-driven motor proteins like myosins, kinesins and dyneins, which move cargos of different types and sizes along the cytoskeleton \cite{albertsbook2013}.

Cytoskeletal transport is an extremely complex process, involving multiple players, long trajectories of the cargos, and a crowded environment. Crowding is significant in two ways.
On one hand, the presence of cytoskeletal fibres is overwhelming: their total length can exceed one metre, huge compared to the typical cell size, of the order of a few tens of micrometres \cite{luby1999}. This implies that the fibres are highly intermingled, with an important number of connections, which can either consist in simple proximity between filaments or be stabilised, or even induced, by accessory protein complexes ({\it e.g.} by Arp2/3 for actin filaments \cite{mullins1999} or augmin complexes for microtubules \cite{goshima2008, petry2013}).
On the other hand, crowding also arises on the strands of filaments themselves, along which motor proteins drag their cargos. This implies that they necessarily interact with one another, if only sterically, and transport by cytoskeletal motor proteins is therefore necessarily a collective process.

It is a fascinating question to ask how the cell-wide delivery can be organised efficiently under these circumstances. In addition, the question of routing arises: depending on the nature of the cargo, delivery will be required to one part of the cell or to another. Even before considering the question how this can be regulated on a cell-wide level, this immediately raises the question of re-routing, {\it i.e.} how motor proteins may be re-directed from one cytoskeletal filament to another, typically at a branching point or crossing of fibres.
 In  some cases, the actors and mechanisms governing track changes have been investigated and are known, such as the action of protein complexes ({\it e.g.} p150 in dynactin cargo complexes for dyneins \cite{mckenney2014}), post-translational modifications of tubuline ({\it e.g.} polyglutamylation \cite{spiliotis2008, janke2011}) and the activity of microtubule-associated proteins (MAPs) \cite{cai2009, atherton2013}.
 However, more generic biophysical mechanisms must be expected to intervene, such as the 3-dimensional proximity of the linear filaments, their elasticity, or  the notion of how molecular motors explore the space around them to find their next point of attachment. To some extent this suggests a contribution of physically motivated, generic rules, that govern the cytoskeletal transport through filament crossings.
 Despite the fundamental progress in microscopy, as well as the advent of single molecule and super-resolution microscopy techniques, it is still a challenge to uncover such generic rules.

Significant advances have nevertheless been made in many respects. Recent literature describes detailed transport features for specific cases, like switching of protofilaments induced by roadblocks \cite{schneider2015} and motion at filament crossings for single cargos driven by one or several motors \cite{ali2007, ross2008, balint2013, miedema2017}.
Employing multiple motors for transporting a cargo constitutes in fact an efficient strategy to overcome a crossing:  due to the fact that motors are distributed over the cargo surface (several hundreds of $nm^2$ for a small organelle) they can explore the 3-dimensional space around the crossing \cite{erickson2011, verdeny2017, bergman2018, gao2018}. In such conditions, research has focussed on the mechanisms that could explain bidirectional cargo transport and regulation of motor activity. Other publications have analysed the transport of a single motor protein {\it in-vitro} and {\it in-cellulo}, and have quantified the statistics of different events which can occur at a crossing: persistent movement across the crossing, switching to another protofilament, reversing, etc \cite{schneider2015, ross2008, balint2013}.  However, little is known when these transport phenomena occur in crowded conditions, meaning when many cargos are occupying the same filament track.    

If cargos are transported by a single motor, the connectivity of filaments and the proximity of potential binding sites on nearby filaments are also expected to matter. Firstly, potential binding sites must be within reach for the next step, and therefore proximity is crucial. But secondly, more subtly, different options for the next processive step will be explored stochastically, weighted according to the free energy landscape of the detailed process of associated conformational changes. This amounts to attributing rates for steps to any of the potential next anchoring points, and sets the search time for the next anchoring position. But how does this impact the global, collective transport of motors through the crossing?

Answering this question experimentally poses huge technical challenges, which reside in  the specific requirements: resolving details of the transport process at a crossing necessitates both very good 3-dimensional resolution, but also  high time resolution. 
For example, the typical step length (from a few tens of $nm$ for dyneins or processive myosins, down to $8~nm$ for  kinesins) sets the length scale which should ideally be resolved. A time scale can be estimated from  motor speeds (between $100~nm/s$ and few $\mu m/s$), which implies that steps and binding/unbinding events occur on the milli- or even micro-second scale. Therefore, the extent of the experimental feat becomes clear: characterising how one motor protein explores the crossing between two adjacent filaments would require to dynamically detect a fully 3-dimensional motor position, with a spatial resolution below $10~nm$ and a time resolution with milli- or micro-seconds.
Imaging microscopy has recently been performed with nanometric lateral resolution \cite{schneider2015, feng2018}, which has indeed provided measurements of the characteristics of various motor protein transport mechanisms. However, time-dependent measurements resolving the motor position in all spatial directions are currently still out of reach for even the most advanced experimental techniques \cite{hancock18, andrecka2018}. In addition, a large majority of imaging methods are based on Single Particle Tracking (SPT) techniques, meaning that collective effects along the cytoskeletal network cannot be probed.

Here we employ a modelling strategy in order to explore the impact of microscopical detail on large-scale transport, including what this implies for experimental measurements.
We acknowledge that a detailed microscopic characterisation of how motors evolve at crossing microtubules is currently unavailable. A description can therefore only be based on effective transition rates, and we elaborate a reduced model, based on stochastic transport on the individual protofilaments. However, we then raise the question of collective transport  as seen through the lens of current experimental options:
all measurable quantities  will characterise the compound microtubule, thus inevitably convoluting them over individual protofilaments. The key question is to which extent theoretical understanding, which is available on the transport process on a single protofilament, may be transposed to  the compound, average transport. To our knowledge such a description has not yet been discussed in the literature. We shall show that this point of view holds surprises, which must be kept in mind when interpreting data on any such measurements.

The paper intends to elaborate a  minimalistic analysis of transport through a crossing of two filaments, exemplified by two nearby microtubules. The model is highly schematic in order to emphasise generic properties, but the framework presented here can easily be extended to more complex situations. The manuscript is organised as follows:
In section II we elaborate a (schematised) representation of a crossing between two nearby microtubules, and we decompose the transport process into single protofilaments with transition points between them.
In section III we briefly review the Totally Asymmetric Simple Exclusion Process (TASEP). The TASEP belongs to the class of exclusion process models that have been successfully applied to cytoskeletal transport driven by motor proteins \cite{parmeggiani2003, parmeggiani2004, leduc2012, miedema2017}, and many theoretical aspects of its collective behaviour are well studied, including on branched structures \cite{embley2008, embley2009, neri2011, neri2013, neri2013b, raguin2013, parmeggiani2014}.
In section III  we employ this paradigmatic model for stochastic and collective transport in order to illustrate our approach as we analyse TASEP transport on our representation of the crossing. We introduce the notion of a 'compound' system, englobing all transport throughout the microtubules without resolving individual protofilaments. This is designed to characterise the overall transport based on coarse-grained, averaged quantities as they may be accessible in experiments. Based on well-established results for TASEP transport on branched structures we establish a compound phase diagram as well as a compound current-density relation. We highlight new features, which qualitatively differentiate the compound system from a branching point between two TASEP segments.  We discuss how these features may interfere in the interpretation of experimental measurements. In particular, the compound current-density relation is seen to be counter-intuitive in two important ways: it no longer constitutes a one-to-one relation between current and density, and it differs according to whether the measurement is made upstream or downstream from the crossing.
In section IV we summarise our results and discuss how the approach employed here can be generalised both to more complex structures and to more complex transport processes. Specifically, we examine how the presence of individual protofilaments is expected to impact predictions for the behaviour of motors as they negotiate a crossing, such as characterised by the recent experimental work of Deeb et al. \cite{deeb2019}. In particular, the traffic through junctions is seen to be intrinsically dependent on crowding, a purely physical effect which will interfere with any active biochemical regulation.
%
The appendices provide a brief review of the mean-field strategy we use throughout, as well as a template for a paper mock-up which may be useful for visualising the transport paths referred to in text and figures.

\section{Unwrapping a microtubule crossing}

\begin{figure*}
  \centering
  \subfigure[]{
    \includegraphics[width=0.35\textwidth]{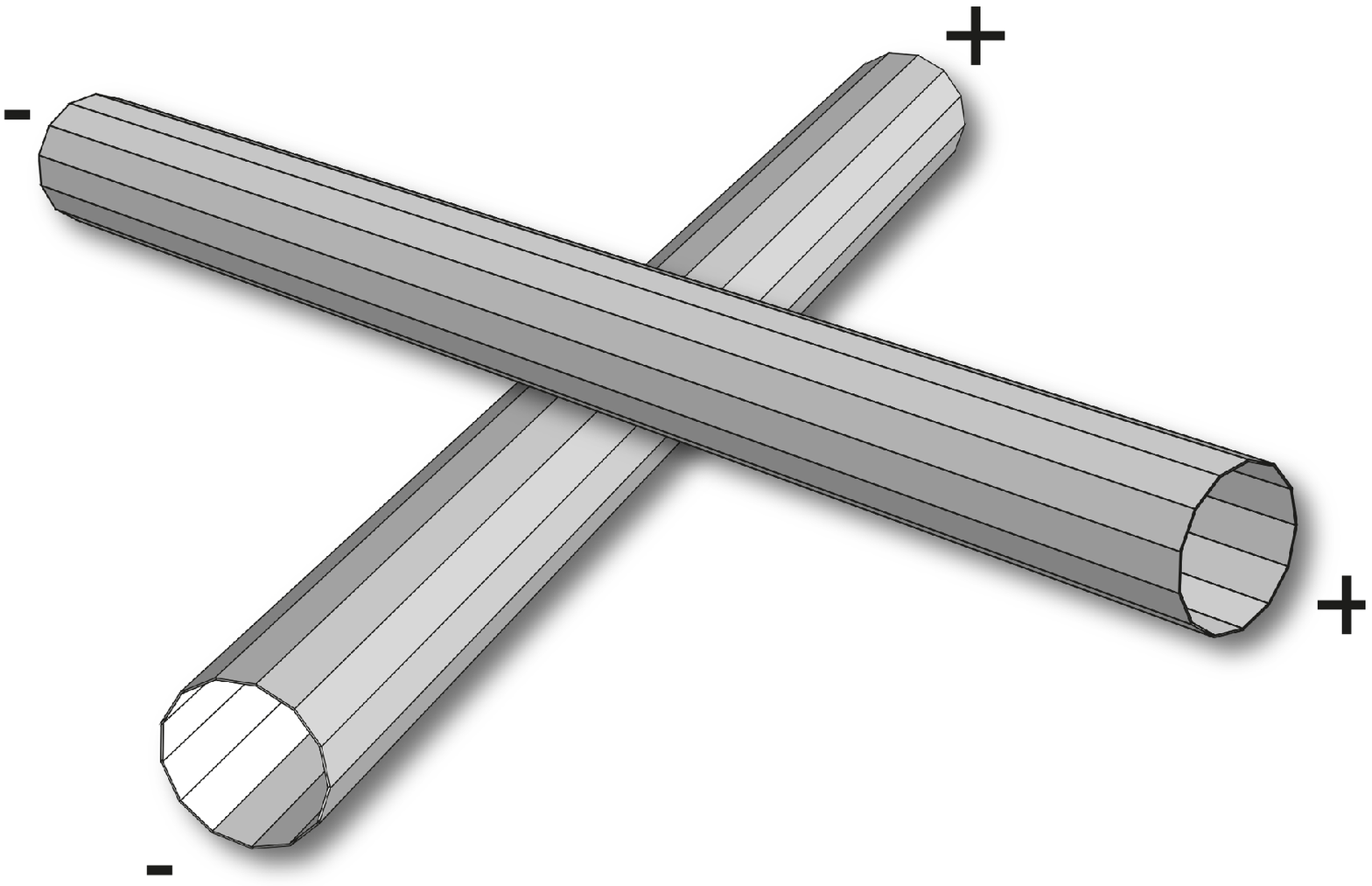}
    \label{fig:crossing:straight}
  }
  \subfigure[]{
    \includegraphics[width=0.6\textwidth]{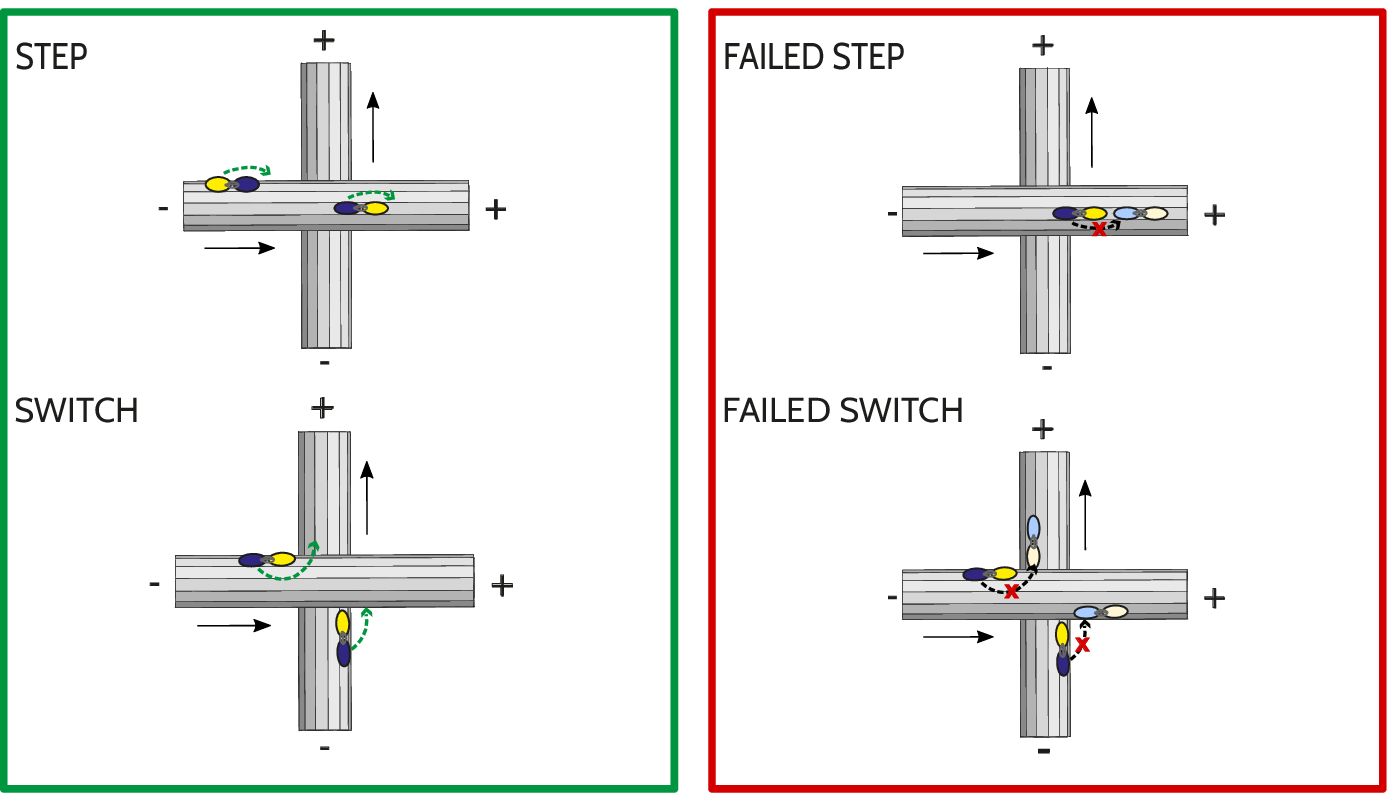}
    \label{fig:crossing:molecules}
    }
  \captionof{figure}{
    (colour online)
    (a) Illustration of a 'crossing', where two microtubules, consisting of (typically) 13 protofilaments each, are in close proximity.  Motor proteins advance stochastically in the direction set by the polarity of each microtubule. As it negotiates the crossing, a motor protein may pause, and will then ultimately follow  straight through on the same microtubule, switch over to the other microtubule or, less frequently, detach.
    (b) Molecular motors require free binding sites ahead in order to advance. The success of the  stepping motion along the microtubules, as well as  switching to the other microtubule, due to obstruction which varies with the density: these are collective effects.
    \label{full microtubules crossing}
  }
\end{figure*}

Traffic of molecular motors  has successfully been modelled as a stochastic process. At this stage we do not focus on any specifics of this process, but rather provide arguments how to disentangle the trajectories of motor proteins in the vicinity of the crossing in a generic manner. To this end, recall that microtubules typically consist of 13 protofilaments, as schematically represented in Fig. \ref{fig:crossing:straight}. The processive motion of a given motor predominantly follows one such filament as a 'track' or 'lane', although 'lane-changes' have been observed \cite{schneider2015, feng2018}. On a larger scale, motors travel along the cytoskeleton, of which the microtubule network is one constituent. Re-routing can occur at 'crossings', where two microtubules are sufficiently close for motors to switch over from one microtubule to another.
They have indeed been seen experimentally to either follow along their initial microtubule, or to switch to the alternative one, as schematically represented in Fig. \ref{fig:crossing:molecules} (a certain number of them may also detach as they approach the crossing, but we do not retain this feature in this work). Here we focus on modelling a crossing of two microtubules, but the strategy is generic, and may be adapted to other situations or other filaments.

To construct our approach it may be useful to have in mind very recent experimental work, such as the results presented by Deeb et al. \cite{deeb2019}. These experiments investigate the way in which motor proteins navigate crossings of microtubules.
Based on single-molecule quantum dot motility assays, the ability of KIF3AC motors to successfully pass crossings of microtubules is studied.
Specifically, probabilities are characterised for motors to either follow straight through a crossing on the same microtubule, or to switch over from one microtubule to the other. Events of motors detaching at the crossing are accounted for (but are relatively rare for KIF3AC). However, the data also quantify the probability for motors to pause before continuing their journey through the crossing. The particular motivation for these experiments has been to discriminate which domains of the motor proteins are involved in successfully transiting the crossing.  By contrasting data for the heterodimer KIF3AC to those on the homodimeric  variants KIF3AA and KIF3CC the study concludes that the KIF3A domain is key to successfully passing such crossings without overwhelming detachment rates.

It is worth pointing out that these experiments are state of the art, and exploit technically challenging imaging techniques in order to detect and characterise the individual events of motors negotiating the crossing. But they also show the current limitations of experimentation.
First, the observation is carried out on single motors. This is of course desired, as it is the most straightforward regime to analyse: as motors are also sparse along the microtubules, no crowding effects can interfere. However, as argued above, this is not necessarily the case in the crowded environment of a cell, and a denser occupancy with motors has indeed been shown {\it in vitro} to lead to important collective effects \cite{leduc2012}.
Second, protofilaments are not resolved in this analysis. This is so for a good reason, which is that motor proteins simply cannot be attributed to specific protofilaments with the current spatio-temporal resolution. It is also perfectly consistent with the setup as single-molecule experiments, for which interactions between motors running on the same protofilament are not an issue.

Both points, however, must be expected to matter for higher motor concentrations: collective effects occur, and especially so between motors following the same protofilament, and it is therefore important to know on which of the parallel protofilaments the motors run. Here we set out to explore the consequences of such a refined description, well ahead of experimental data, in order to highlight that this leads to qualitatively new phenomena as well as unexpected pitfalls. We will also show specifically that the probabilities for 'straight' or 'switch' events at the crossing are expected to evolve with the density of motors.
In our analysis we thus remain within the picture of motors following specific tracks, but which we take to be individual protofilaments. This implies that one must then also ask which paths the motors take, or may take, when transitioning from one microtubule to another: this is the extra complexity which, as we will show, gives rise to new features.

We first focus on analysing a (deliberately) schematic model for this process, on a simplified representation of a microtubule crossing, which serves to acknowledge the effect of this additional complexity.
Unravelling all possible paths along protofilaments on a simplified microtubule, we then argue that these correspond to a network of interlinked tracks, which can be analysed using the established methods for transport on branched structures. We then show how to reconstruct transport through the microtubule crossing, considered as a compound of possible paths along protofilaments.
This prepares the ground for analysing the phases and the current-density relation of the compound crossing, as it would be observable: this will reveal significant differences from what one would expect based on intuition of single-track models.

We conclude by discussing how this approach may be extended to more refined, less schematic representations of microtubules, illustrating also how switch rates become density dependent due to collective effects in the transport process.

\subsection{Protofilaments and transport along elementary paths}\label{Modelisation of the crossing}

\begin{figure*}
  \centering
  \subfigure[]{
    \includegraphics[width=0.3\textwidth]{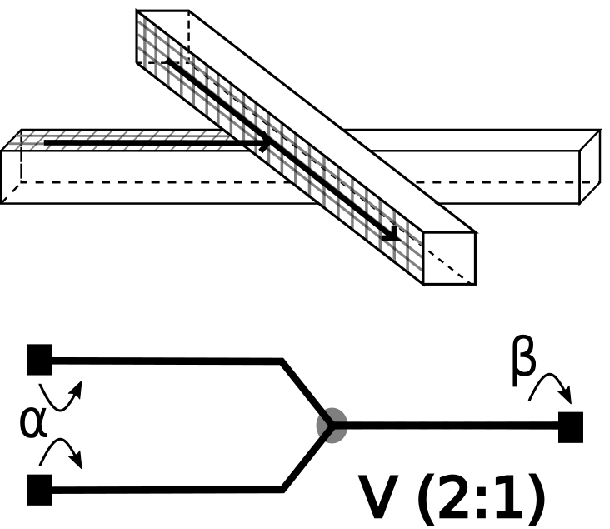}
  }
  \subfigure[]{
    \includegraphics[width=0.3\textwidth]{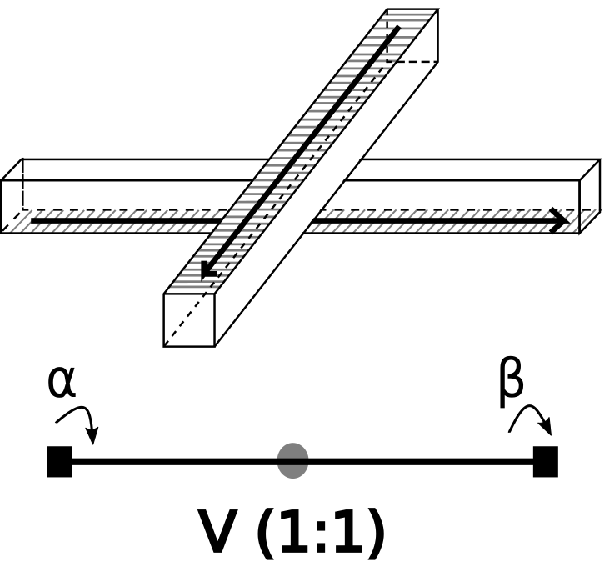}
  }
  \subfigure[]{
    \includegraphics[width=0.3\textwidth]{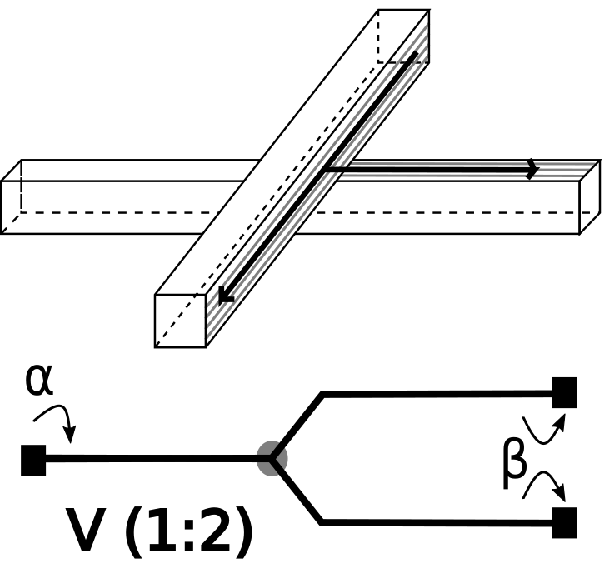}
  }
  
  \captionof{figure}{
    Schematic representing crossing microtubules by two parallelepipeds, each face representing a protofilament. We assume that motor proteins can only switch from one track to another if the latter shares a common edge.
    Motor proteins can only follow three kinds of paths along the lattice, called vertices: (a) V(2:1) (two entries, one exit at the branching point), (b) V(1:1) (two independent, unbranched trajectories) and (c)  V(1:2) (one entry, two exits at the branching point).
  A mockup to  be cut out from paper is supplied in Appendix \ref{app:mockup}, and may prove useful for visualising the proximity of tracks.
  } 
  \label{microtubules crossing}     	
\end{figure*}

On each protofilament, the flow of motor proteins  can be described by a current-density relation of the form
\begin{equation}
  \label{eq:tasep:Jrho}
  J=  \rho \, v(\rho)
  \ ,
\end{equation}
where $J$ is the current (number of motors passing a given point per unit time), $\rho$ is the motor density (number per unit length), and $v$ is their  (average) velocity. This is often referred to as the {\it fundamental relation}, and is central to phenomena of collective transport \cite{schadschneider2000}. The dependency $v(\rho)$ captures all collective effects due to the interaction of motors, such as steric hindrance or more complex phenomena. This relation may be derived from the stochastic transport process under consideration, or $v(\rho)$ can be a phenomenological relation, deduced from experimental data. 
The point is that such relations are available for molecular motors \cite{leduc2012,miedema2017}, restricted to a single filament, {\it i.e.} in the absence of branching. Here we show that decomposing the crossing of microtubules into interconnected paths makes it possible to study this problem for a compound, multi-track crossing.

The idea is to analyse the paths along individual protofilaments which motor proteins may take as they travel along the microtubule. This will unwrap the 3-dimensional structure of the crossing into an ensemble of 1-dimensional or quasi-1-dimensional paths, corresponding to the protofilaments, with branching points located at the crossing. We will then proceed, in the next section, to solve the collective transport on this ensemble of filaments for a particular transport model and illustrate the relevance of our findings in terms of what to expect from finite resolution measurements.

In our model we need to acknowledge a certain number of facts: first, as   motor proteins approach a crossing on a given protofilament, they may simply follow their path, presumably along the same protofilament; but they may also transition to the crossing microtubule. The trajectory for any given motor protein is a probabilistic process, but their likelihoods must be expected to vary from one protofilament to the other: indeed, due to the cylindrical structure of the microtubules, those protofilaments located on the side facing the second microtubule may offer the possibility for a motor protein to transition, whereas this is much less likely for the protofilaments on the opposite side. Steric hindrance due to the obstruction presented by the crossing microtubule is another feature expected to influence the motion of a motor protein, which may be seen as a variant of  the 'roadblocks' considered in \cite{schneider2015}.
\\
Here we account for this complexity in a deliberately schematic way, by distinguishing three types of trajectories along a protofilament  (see Figure \ref{microtubules crossing} for an illustration)~:
  (i) on certain protofilaments, close to the crossing microtubule, a motor protein cannot pursue its trajectory and is obliged to transition to the closest alternate protofilament;
  (ii)  on the opposite side of the microtubule, a motor protein follows its trajectory on the same protofilament, without transitioning, as there is no specific reason to do so; 
  (iii) in intermediate positions, the motor protein randomly does one or the other; for simplicity, we assume a probability of 50\% for each option, although this is a parameter which should be investigated further.
  \\
  The simplest setup retaining these scenarios requires 4 protofilaments for each microtubule. Although this is of course different from real microtubules, we make this simplifying choice as it allows us to focus on qualitative arguments: it will be obvious in the discussion that a generalisation to 13 protofilaments (or other protofilament configurations) is straightforward. In order to visualise our representation of the crossing between two microtubules, we refer to Figure \ref{microtubules crossing}, where the four protofilaments are represented as the geometrical sides of a square tube (parallelepiped).
  We provide an A4 scaled mock-up (see Appendix \ref{app:mockup}) for easier visualisation.
In this representation transitions from a protofilament to another one on the crossing microtubule are therefore possible between faces which share an edge, and compulsory if the face is obstructed by the crossing microtubule.

Figure \ref{microtubules crossing} also indicates the topology of the trajectories as they unwrap along protofilaments. Several types of paths of motor proteins along protofilament tracks can be identified, as is illustrated in Figure \ref{microtubules crossing}: subfigures (a), (b) and (c) correspond to scenarios (i), (ii) and (iii) above, respectively. The schematic diagrams below the illustrations identify the corresponding protofilaments as tracks joined at a contact point, labeled as a V(2:1) where two tracks join at the crossing, as V(1:1) for the tracks without branching, and as V(1:2) where two exiting tracks are available for the molecular motor (the nomenclature follows our earlier work, see \cite{embley2009}). We consider that, far away from the crossing, all protofilaments are in contact with reservoirs, represented by an entry rate $\alpha$ and an exit rate $\beta$, which are common to all protofilaments.

\subsection{Averaging and experimental resolution\label{sec:averages}}

The key lies in the observation that, as argued in the introduction, the spatial resolution in measurements must be accounted for: transport on individual protofilaments cannot currently be resolved in crowded environments, and we are therefore necessarily dealing with an average over parallel protofilaments. Noting $i=1...P$ the individual protofilaments making up the microtubule, the average density ($\rho$) and the average current ($J$) are given by
\begin{equation}
  \label{eq:averaging}
 \begin{array}{llllllll}
\rho(x) = \frac{1}{P} \sum\limits_{i=1}^{P}\rho_{i}(x) \ \ \ \mbox{and} \ \ \ J(x) = \frac{1}{P} \sum\limits_{i=1}^{P}J_{i}(x)\ . 
\end{array}
\end{equation}

Here $\rho_i(x)$ is the motor density on protofilament $i$, which may depend on the spatial position $x$ measured along the protofilament. The total current on the other hand will be independent of the position, due to conservation, unless attachment/detachment of motors is considered.
We will refer to  the {\it compound} current $J$ and density $\rho$, to distinguish them from those of individual protofilaments $J_i$ and $\rho_i$. Experimental observation would in fact measure $P\times J(x)$ and $P\times \rho(x)$.

 Up to this point the model of the crossing is not specific to any particular microscopic representation of the transport process, and it is only in the following section that we will use the paradigmatic TASEP as one model to illustrate the important implications this analysis has for measurements on the compound microtubule crossing.

\section{TASEP transport through a microtubule crossing}

In this section we exploit the model we have developed for the crossing of two microtubules and explore the resulting phenomenology, based on one particular choice for representing the collective transport, the Totally Asymmetric Simple Exclusion Process (TASEP). We briefly summarise the main features of the model and recall how it can be adapted to treat branched paths in a mean-field approximation. We establish the features of a 3-fold branching point. From this we construct the full transport characteristics through the compound microtubule crossing, and discuss the resulting phenomenology.

\subsection{A brief reminder of TASEP transport}\label{Biref reminder}

We base our analysis on the TASEP model, a well-studied model for stochastic but directed transport on a line-like structure, in which interactions between particles are limited to an excluded volume condition.  Any such TASEP segment consists of successive sites, corresponding to the potential positions of molecular motors on the periodic protofilament structure, which may be empty or occupied at any instant. Time-averaging the occupancy for each site yields a local density for each site, and thus a spatial density profile. Motor proteins stochastically advance one site, with rate $\gamma$, if and only if the target site is not yet occupied (excluded volume interaction). The boundary conditions consist in attempting to inject a particle on the first site (entrance rate $\alpha$) or removing one from the last site (exit rate $\beta$), which amounts to coupling the segment to two particle reservoirs. The model is summarised graphically in Figure \ref{TASEP_1segment}.  An exact solution exists \cite{derrida1993, schutz1993}, and validates the simple mean-field expression which can be established as follows.

\begin{Figure}
  \centering	
  \includegraphics[width=0.75\textwidth]{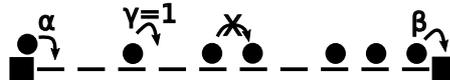}      
  \captionof{figure}{
    Schematic illustration of a TASEP lattice with open boundary conditions. The attempt rate for a particle to enter is $\alpha$, and the exit rate $\beta$. Particles hop from left to right, and interact through their excluded volume. We use $\gamma=1$ throughout.
    }
  \label{TASEP_1segment}  
\end{Figure}

\subsubsection{Collective transport in TASEP} \label{Collective transport in TASEP}

The behaviour of this model is well studied \cite{chou2011}. One central result is that, with the exception of small  zones at the boundaries  which can be interpreted as boundary layers from a hydrodynamic perspective, the density profile is homogeneous throughout the entire segment \cite{schutz1993}. Denoting this bulk density $\rho$ it directly follows that the mean-field current is given by 
\begin{equation}
  J
  = \gamma \, \rho(1-\rho)
  \ .
\label{current_density}
\end{equation}
Here $\gamma$ is the  microscopic jump rate. It is commonly taken to be equal to unity, which amounts to using it to set the timescale.

\begin{figure*}
  \centering
  \subfigure[]{
    \label{TASEP 1D}
    \includegraphics[width=0.5\textwidth]{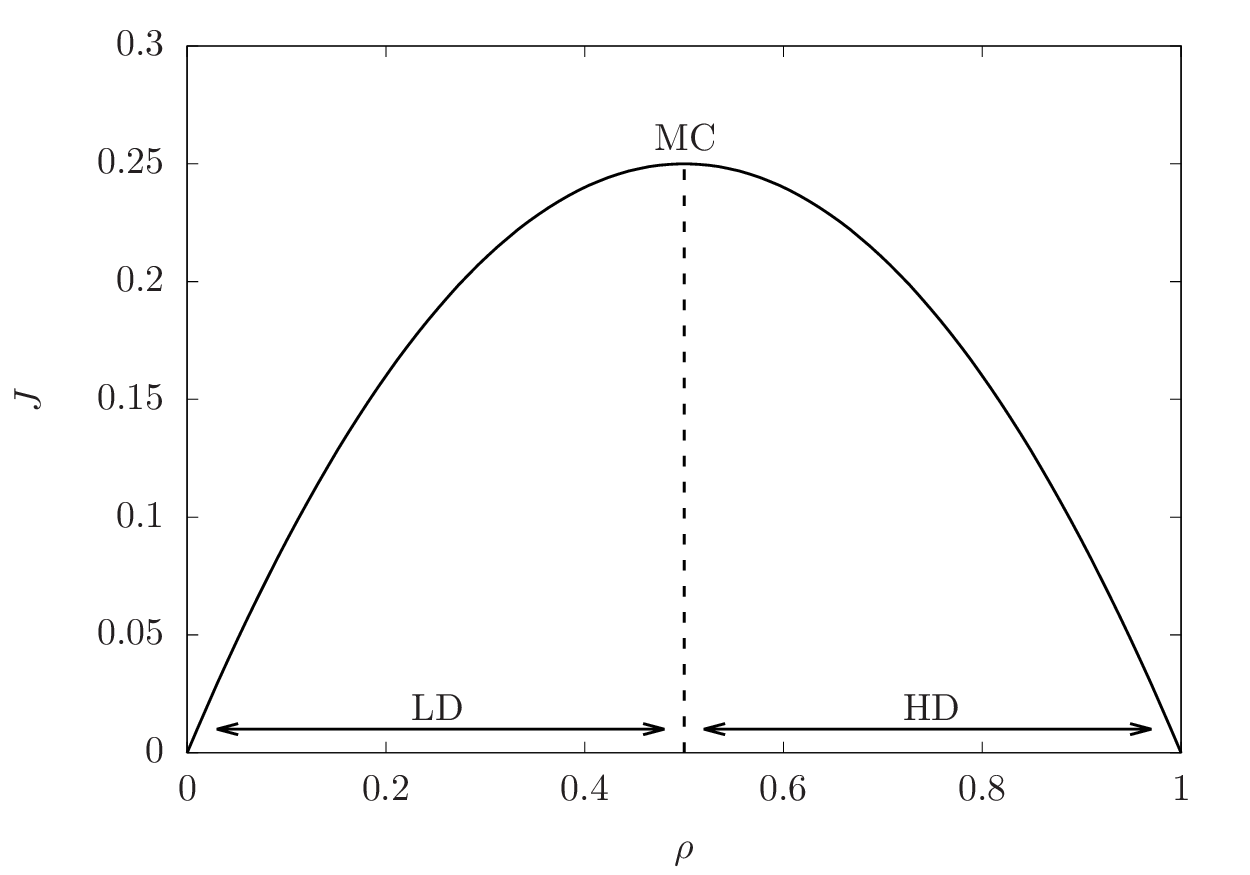}
  }
  \hspace{1.0cm}
  \subfigure[]{
    \label{parabole}
    \includegraphics[width=0.3\textwidth]{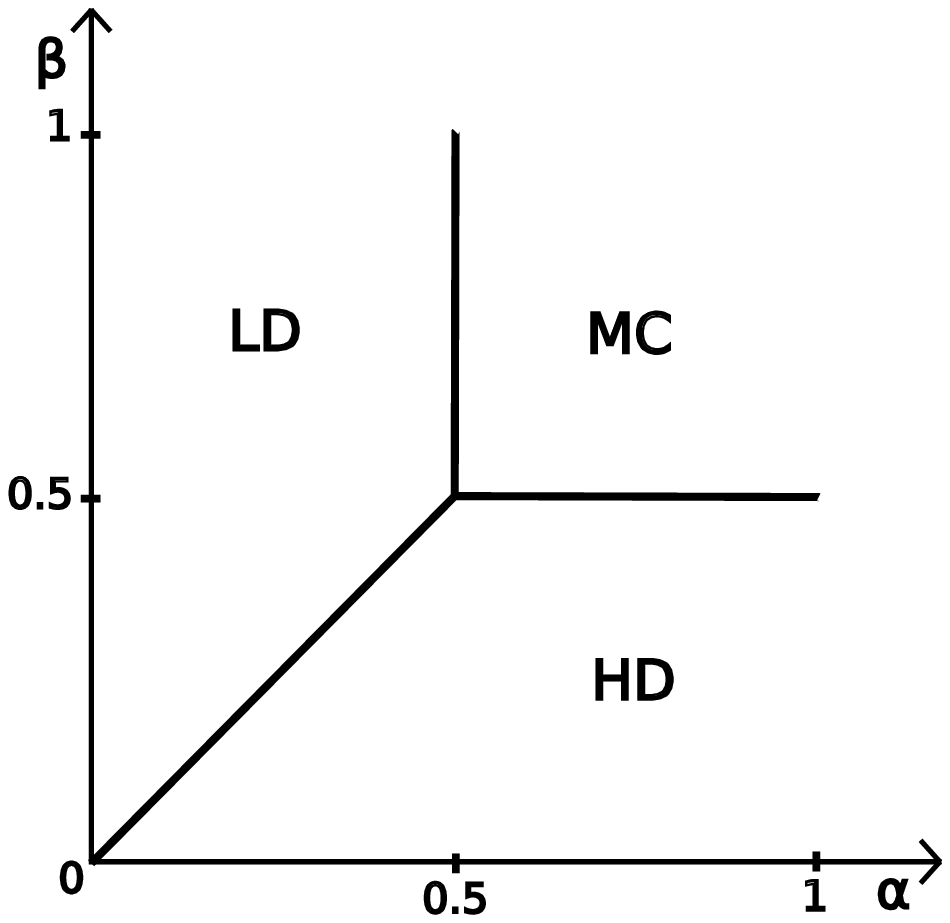}
  }
  \captionof{figure}{
    Main features of TASEP transport.
    (a) fundamental relation of transport for a simple TASEP segment with open boundary conditions, $J=\rho(1-\rho)$.        
    (b) Corresponding phase diagram, featuring one of the three transport regimes according to the entry and exit rates ($\alpha$ and $\beta$):
    in the LD (low density) phase the current is entry limited, in HD (high density) it is exit limited, and in MC (maximum current) it is bulk limited.
  }        	
\end{figure*}

Distinct transport regimes arise, which are controlled by the boundary rates ($\alpha$ and $\beta$). In essence, one observes a low density (LD) phase when the entry rate is limiting ($\alpha<\beta$ and $\alpha<\gamma/2$), a high density (HD) phase when the exit rate is limiting ($\beta<\alpha$ and $\beta<\gamma/2$), and a maximum current (MC) phase when  the bulk hopping rate is limiting ($\gamma/2<\alpha$ and $\gamma/2<\beta$).
The bulk density is given as $\rho=\alpha$ (in LD), as $\rho=1-\beta$ (in HD), and as $\rho=1/2$ (in MC); the corresponding currents follow from Eq. (\ref{current_density}). In the following we set $\gamma=1$, which amounts to measuring all rates in units of $\gamma$.

These regimes can be identified directly from the current-density relation $J(\rho)$ (see Figure \ref{TASEP 1D}), and are summarised in the representation referred to as a 'phase diagram' (Figure \ref{parabole}), which attributes a phase to each point in the $(\alpha,\beta$) plane of parameter space. These phases are key to the following analysis.

\subsubsection{Branched paths and effective rates} \label{section Phase diagram for branched lattices}

As argued above, we have decomposed the possible trajectories a given  motor protein may take along a microtubule crossing into a V(1:1), {\it i.e.} a direct straight path along a protofilament, as well as 3-fold branching points, a V(2:1) where two protofilaments join, and a V(1:2) where a protofilament splits into two possible choices. Features of this system have been studied \cite{kolomeisky2005}, and more complex branched structures have been analysed in great detail using mean-field arguments based on effective rates \cite{embley2008, embley2009, raguin2013}.

\begin{Figure}
  \centering	
  \includegraphics[width=0.75\textwidth]{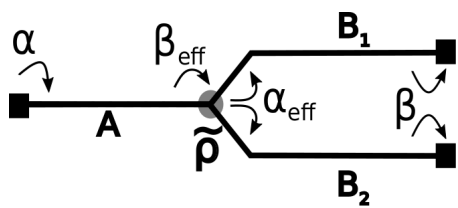}      
  \captionof{figure}{
    Schematic illustration of a vertex V(1:2) with open boundary conditions characterized by and entry rate $\alpha$ and an exit rate $\beta$. At the branching point the density is $\tilde{\rho}$, and the effective rates are $\beta_{eff}$ (to exit segment A) and $\alpha_{eff}$ (to enter either of segments B).
    \label{rates_V12}
  }
\end{Figure}

To summarise briefly, the idea behind effective rates is to focus on the site at the point where two paths meet. This is considered simultaneously as an exit reservoir to the upstream segment(s) and as an entry reservoir to the downstream segment(s), as represented in Figure \ref{rates_V12}. This amounts to decomposing the branched paths into simple TASEP segments, coupled at the branching points.

\subsubsection{TASEP with 3-fold branching points\label{sec:branchedTASEP}}

These arguments can directly be applied to the 3-fold vertices.
 It is straightforward to see \cite{embley2009} that in a mean-field spirit, noting the density at the branching site $\tilde\rho$, we have effective rates of
\begin{equation}
  \beta_{eff} = 1-\tilde\rho, \qquad \alpha_{eff}=\tilde{\rho}/2 
\end{equation}
for a V(1:2), and
\begin{equation}
  \beta_{eff} = 1-\tilde\rho, \qquad \alpha_{eff}=\tilde{\rho}
\end{equation}
for a V(2:1). See also the brief discussion in Appendix \ref{Appendix on 3-fold vertices}.

Current conservation then fixes the density $\tilde\rho$, from which everything else follows. Details of these calculations are not essential for the interpretation, and we recapitulate the method in Appendix \ref{Appendix on 3-fold vertices}.
Here we simply summarise the results as they are relevant for the following arguments.

\begin{figure*}
  \centering
  \subfigure[]{
    \includegraphics[width=0.3\textwidth]{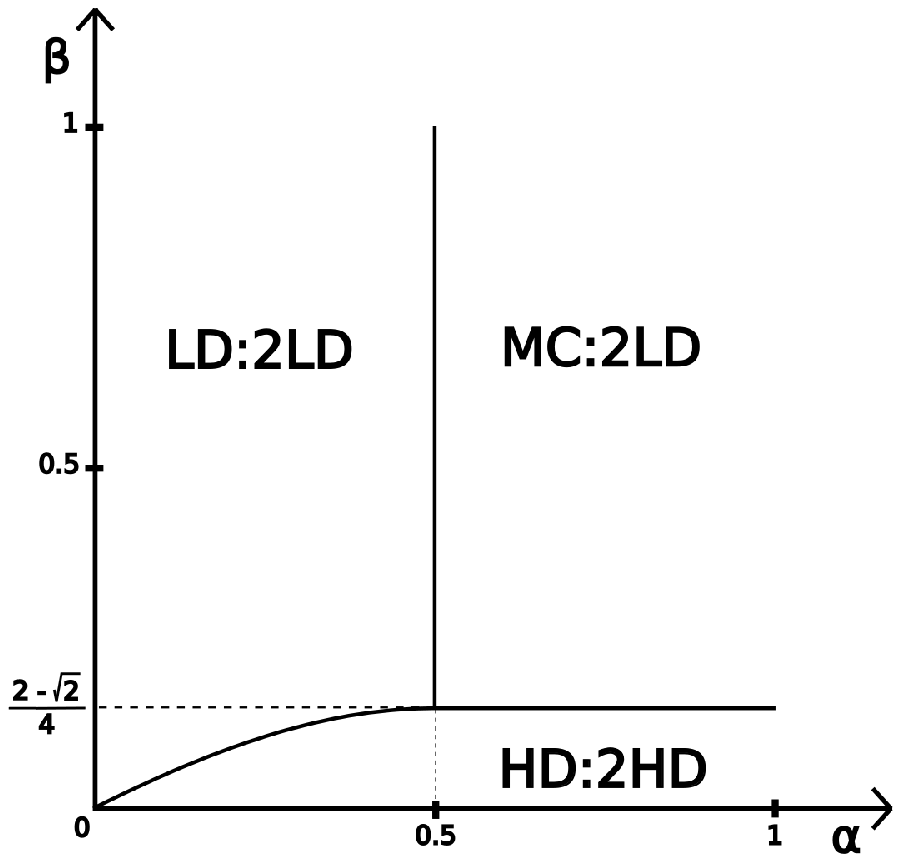}
    \label{phase_diagram_V12}
  }
  \hspace{2.5cm}
  \subfigure[]{
    \includegraphics[width=0.3\textwidth]{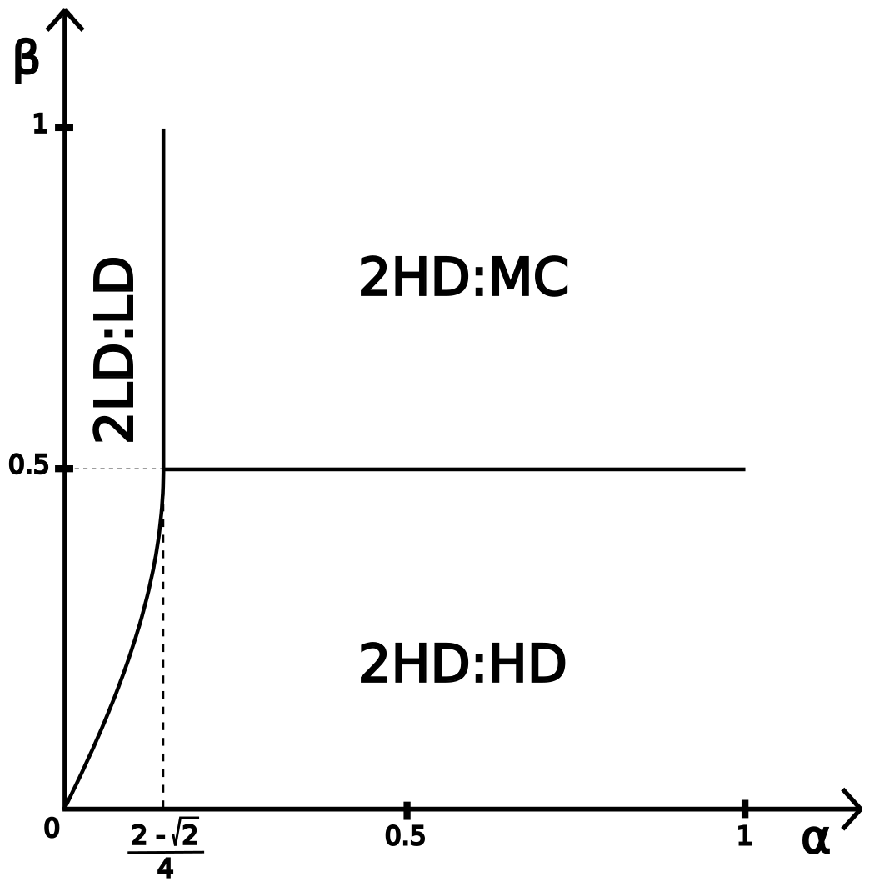}
    \label{phase_diagram_V21}
	}
  \captionof{figure}{
    Phase diagrams of 3-fold vertices (a) V(1:2) and (b) V(2:1) with open boundary conditions. The exit rates for the V(1:2) are taken to be identical for both B-segments, just as the entrance rates for the V(2:1) are identical for both A-segments. Both diagrams are obviously related, as can be argued based on an extended particle-hole symmetry for branched structures, analogous to what has been discussed for a similar system \cite{embley2008, embley2009}.
	}  
\label{V12_V21_phase_diagrams}      	
\end{figure*}

Following the notation used in \cite{embley2009}, we characterise the state of the branched system by indicating the phases in each of its segments as A{:}(B$_1$|B$_2$) for a V(1{:}2) or as (A$_1$|A$_2$){:}B for a V(2{:}1). As we are assuming that both out-segments are equivalent in a V(1:2), and both in-segments in a V(2:1), the notation can be simplified to A:2B in the former case and 2A:B in the latter.
The main results are the phase diagrams, in which we find the 3 TASEP phases. With respect to a simple TASEP segment though, the critical rates at which the transition into the MC phase occurs are modified, and the phase boundary separating LD and HD phases is no longer a straight line.
Analytical mean-field expressions for the various phases are recapitulated in Table \ref{table:vertices} for the density, the current can be deduced directly by applying Eq. (\ref{current_density}) to the segment which carries the full current. 

It will be useful to recall the correspondence between the two phase diagrams in Figure \ref{V12_V21_phase_diagrams}, which can be established on the grounds laid out in \cite{embley2009}. Essentially, it consists in realising that a V(1:2) corresponds to a V(2:1) operated 'backwards', and vice versa. This requires reverting the flow direction, and therefore exchanging the entrance rate $\alpha$ and the exit rate $\beta$. Those segments on which transport was limited by the entrance rate (LD) therefore become limited by the exit rate (HD), and vice versa, whereas those limited by the bulk flow (MC) remain unchanged.  Applying all these changes simultaneously maps the two phase diagrams onto each other. This correspondence can be linked to a particle-hole symmetry in the TASEP model \cite{embley2009}.

\renewcommand{\arraystretch}{1.2}
\newcolumntype{P}[1]{>{\centering\arraybackslash}p{#1}}  
\begin{center}
  \begin{table*}[htbp]
    \centering
    \caption{
      Mean-field expressions characterising all phases, for both 3-fold vertices, V(1:2) and V(2:1), respectively. For each phase (column 1), the corresponding conditions on entrance and exit rates are specified (columns 2 and 3). Densities on the sections upstream ($\rho_A$) and downstream ($\rho_B$) are indicated (columns 4 and 5), as well as the density $\tilde{\rho}$ on the branching site itself (column 6) and the current through the branching point (column 7). To improve readability, expressions are stated in terms of the currents $J_\alpha=\alpha(1-\alpha)$ and $J_\beta=\beta(1-\beta)$. Note that the densities $\rho_{A,B}$ in the column for the density at the branching point $\tilde{\rho}$ refer to the preceding columns on the same line.
       \label{table:vertices}
    }
    \begin{tabularx}{\linewidth}{%
        P{.12\textwidth} %
        P{.15\textwidth} %
        P{.15\textwidth} %
        P{.11\textwidth} %
        P{.11\textwidth} %
        P{.12\textwidth} %
        P{.10\textwidth} %
      }   
      \hline\hline
      \textbf{phase} & \textbf{interval of $\alpha$} & \textbf{interval of $\beta$} & \textbf{$\rho_A$} & \textbf{$\rho_B$} & \textbf{$\tilde{\rho}$}  & \textbf{$J$}\\
      \hline
      \multicolumn{2}{l}{\textbf{vertex V(1:2)}}\\
      LD:2LD & $\alpha <1/2$ & $\beta> \frac{1-\sqrt{1-2J_\alpha}}{2}$ & $\alpha$ &  $\frac{1-\sqrt{1-2J_\alpha}}{2}$ & $2 \rho_B$ & $J_\alpha$\\
      MC:2LD & $\alpha>1/2$ & $\beta>\frac{2-\sqrt{2}}{4}$ & 1/2 & $\frac{2-\sqrt{2}}{4}$ & $2 \rho_B$ & 1/4\\
      HD:2HD & $\alpha>\frac{1-\sqrt{1-8J_\beta}}{2}$ & $\beta<\frac{2-\sqrt{2}}{4}$ & $\frac{1+\sqrt{1-8J_\beta}}{2}$ & $1-\beta$ & $\rho_A$ & $2J_\beta$\\
      \hline
      \multicolumn{2}{l}{\textbf{vertex V(2:1)}}\\
      2LD:LD & $\alpha <\frac{2-\sqrt{2}}{4}$ & $\beta> \frac{1-\sqrt{1-8J_\alpha}}{2}$ & $\alpha$ &  $\frac{1-\sqrt{1-8J_\alpha}}{2}$ & $\rho_B$ & $2J_\alpha$\\
      2HD:MC & $\alpha>\frac{2-\sqrt{2}}{4}$ & $\beta>1/2$ & $\frac{2+\sqrt{2}}{4}$ & 1/2 & $\rho_A$ & 1/4\\
      2HD:HD & $\alpha>\frac{1-\sqrt{1-2J_\beta}}{2}$ & $\beta<1/2$ & $\frac{1+\sqrt{1-2J_\beta}}{2}$ & $1-\beta$ & $\rho_A$ & $J_\beta$\\
      \hline\hline
    \end{tabularx}
  \end{table*}
\end{center}

\subsection{Constructing transport through the compound microtubule crossing}

In this section we exploit the results discussed above to address the question of compound transport through the three dimensional structure of two crossing microtubules. So far we have broken down the complex pathways along this complex structure in section \ref{Modelisation of the crossing}, and then analysed transport on each, possibly branched, pathway in section \ref{sec:branchedTASEP}. Here we set out to assemble the results, to construct transport through the structure in different regimes of crowding. Recall that measurements do not currently give access to the transport through individual protofilaments, but rather cumulate those into a global current-density relation for the entire microtubule, which implies an average as discussed in section \ref{sec:averages}. Here we derive an expression for this compound current-density relation based on TASEP transport.

\subsubsection{Compound phase diagram \label{eq:compoundPhaseDiagram}}

Assuming that, away from the crossing, all protofilaments connect to the same entry or exit reservoirs, they are all subject to the same entry and exit rates, $\alpha$ and $\beta$. The state of the full system can therefore be characterised by determining the phase of each of the segments in all paths, for each type of vertex, for any point $(\alpha,\beta)$ in the parameter plane. This amounts to superposing the phase diagrams for the branched vertices V(1:2) and V(2:1), as well as the regular segment V(1:1), on the same phase plane. This superposition is represented in Figure \ref{diagramme_superpose}, where each line of the labels refers to one type of vertex.
\\

From this diagram we can read off, for any given set of rates $(\alpha,\beta)$, the phases in all segments of all protofilaments, as represented in the labels.
Based on the expressions in Table \ref{table:vertices} for the individual types of 3-fold vertices, the corresponding densities and currents can be attributed to each of these phases. All in all this fixes the transport through any of the individual protofilaments of the microtubule crossing.

\subsubsection{Symmetry \label{sec:symmetry}}

The symmetry in the compound phase diagram is apparent. It can be understood by realising that the compound transport regime can be established by superposing those of its constituting branched paths:  V(1:2), V(2:1) and V(1:1). Applying the correspondence between topologically complementary  V(1:2) and V(2:1), discussed above (see section \ref{sec:branchedTASEP}), one can deduce how compound phases transform into one another. For example, the compound phase II, characterised as (LD{:}2LD; 2HD{:}MC; LD{:}LD) would map  onto (2HD{:}HD; MC{:}2LD; HD{:}HD). Note however that the order in which the vertices composing the compound structure are enumerated is pure convention, here taken to be the order (V(1:2); V(2:1); V(1:1)). This therefore corresponds to a mapping between the phases II and VI in the compound phase diagram.

\subsubsection{Compound densities and currents}

In order to make contact with experimentally measurable quantities, we now establish expressions for the densities as they might be observed on the compound microtubule. Indeed, as outlined before, we are chiefly interested in the density as it would be probed over a zone of finite size, implying an average both over parallel protofilaments as well as over a finite distance along each filament. The latter average is innocuous, as both density and current are constant along each filament (with the exception of small boundary layers).
\\

\begin{figure*}
  \centering
  \subfigure[]{
    \includegraphics[width=0.45\textwidth]{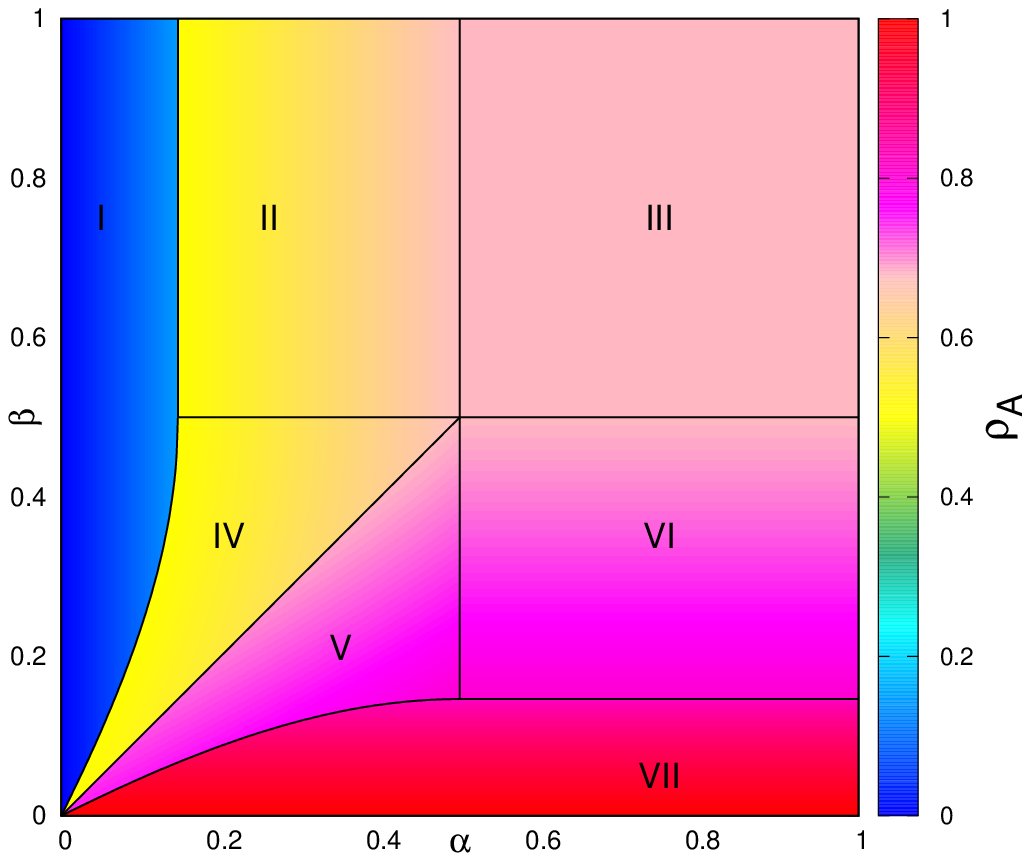}
    \label{heat_map_rho_A}
  }
  \hspace{1.0cm}
  \subfigure[]{
    \includegraphics[width=0.45\textwidth]{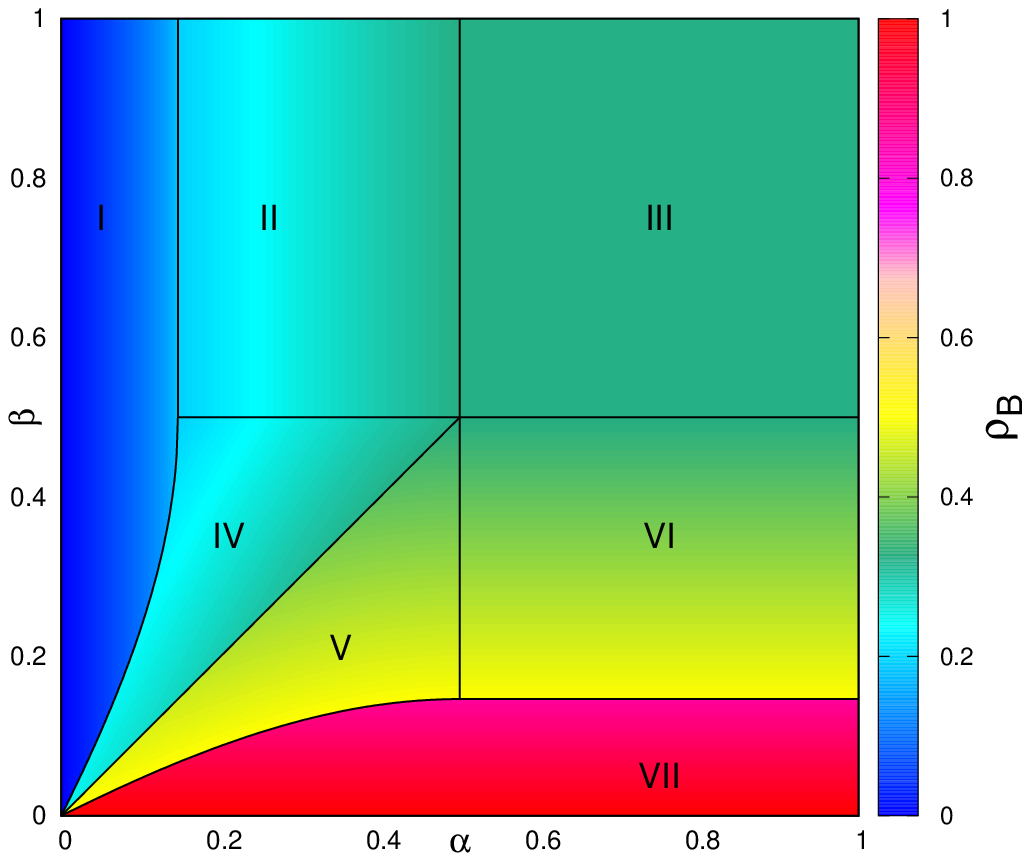}
    \label{heat_map_rho_B}
  }
  \captionof{figure}{
    (colour online) Density, colour-coded for each set of boundary rates $(\alpha,\beta)$, for the full crossing. (a) Upstream density ($\rho_A$) and (b) downstream density ($\rho_B$). The values are obtained numerically, by enumerating the values of these rates and exploiting the mean-field expressions of Table \ref{table:vertices}. The phase boundaries sketched in Figure \ref{diagramme_superpose} above are clearly visible.
    Phases are denoted by capital roman numbers, which refer back to Figure \ref{diagramme_superpose}.
   The difference between panels (a) and (b) stresses how the location of the measurement zone impacts the observed results:  in identical conditions, {\it i.e.} same rates $(\alpha,\beta)$, the {\it compound} density largely differs between the zones upstream versus downstream of the branching point for most phases.
}  \label{heat_maps}      	
\end{figure*}

Nevertheless, we must consider two cases, according to whether the measurement is performed upstream (A segments) or downstream (B segments) of the crossing.  Referring to the mock-up of the crossing provided in Appendix \ref{app:mockup} will illustrate this easily. Performing the averages over protofilaments (see Eq. (\ref{eq:averaging})) yields the density on sections upstream of the crossing as
\begin{equation}
  \label{eq:rhoA}
  \rho_{A}=\frac{1}{4}[2 \,\rho_{A,V(2:1)} + \rho_{A,V(1:2)} + \rho_{A,V(1:1)}]\ ,
\end{equation}
whereas downstream it is
\begin{equation}
  \label{eq:rhoB}
\rho_{B}=\frac{1}{4}[\rho_{B,V(2:1)} + 2 \, \rho_{B,V(1:2)} + \rho_{B,V(1:1)}]\ ,
\end{equation}
with densities specific to each vertex type given as in Table \ref{table:vertices}.
These are the densities which would be observed by a local measurement, such as those based on fluorescence, averaging over the protofilaments due to lack of resolution in crowded conditions.
The resulting compound density, upstream and downstream of the crossing, is represented in Figure \ref{heat_maps}, and the colour-coded values of the density are seen to feature different zones, which clearly reflect both, the structure of the phase diagram for a full crossing, and the symmetry discussed above.

\begin{Figure}
  \centering
  \includegraphics[width=\textwidth]{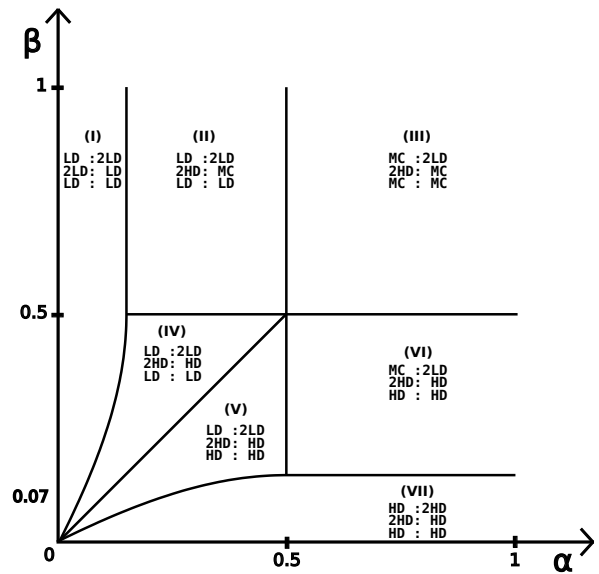} 
  \captionof{figure}{
    Analytically determined phase diagram of the compound model for two crossing microtubules. It is obtained as the superposition of the phase diagrams of vertices V(1:2), V(2:1), and V(1:1). Capital roman numbers are used to identify the individual phases for later reference.
  }
  \label{diagramme_superpose}
\end{Figure}

\subsubsection{Sweeps in the entry rate}

A common strategy in experiments is to control the density of molecular motors in the system by changing their concentration in the surrounding. This amounts to setting the in-rate $\alpha$, in our case for the compound system, whereas the exit rate $\beta$ remains unaffected. Measuring the current for several such values of $\alpha$, while keeping $\beta$ constant, thus yields a graph of $J(\alpha)$, which is a useful way of characterising transport regimes.

\begin{figure*}[htbp]
  \centering
  \, \hspace*{-1.5cm}
  \begin{tabular}{lccc}
    & $\beta=0.07$
    & $\beta=0.25$
    &$\beta=0.75$
    \\
    &\includegraphics[width=0.3\textwidth]{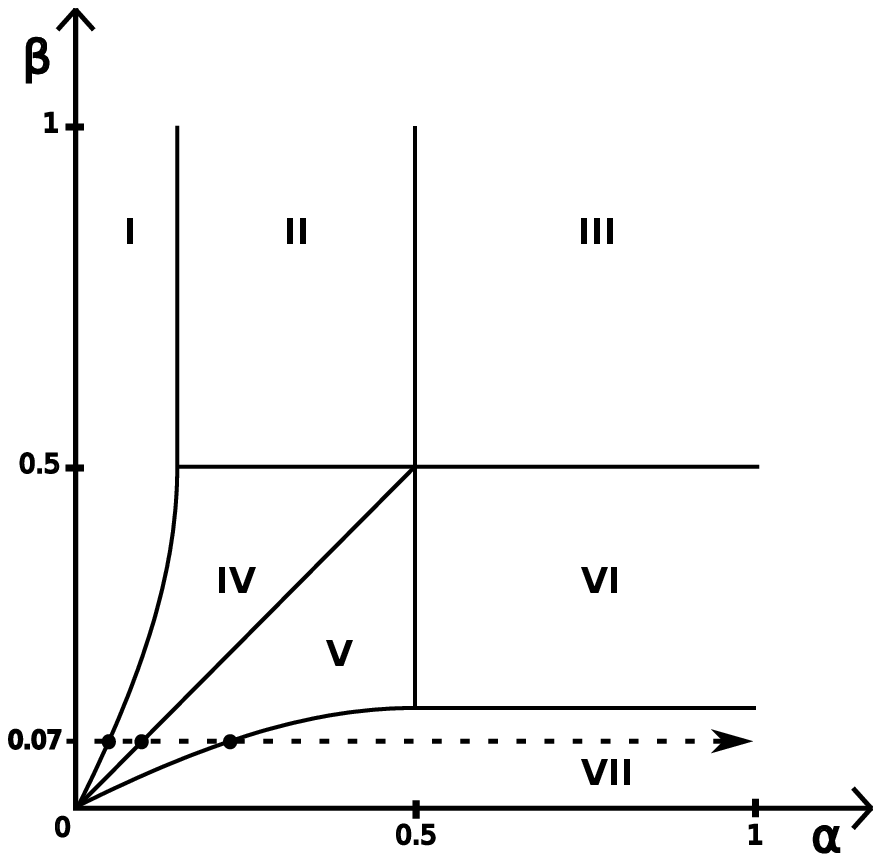}      
    &\includegraphics[width=0.3\textwidth]{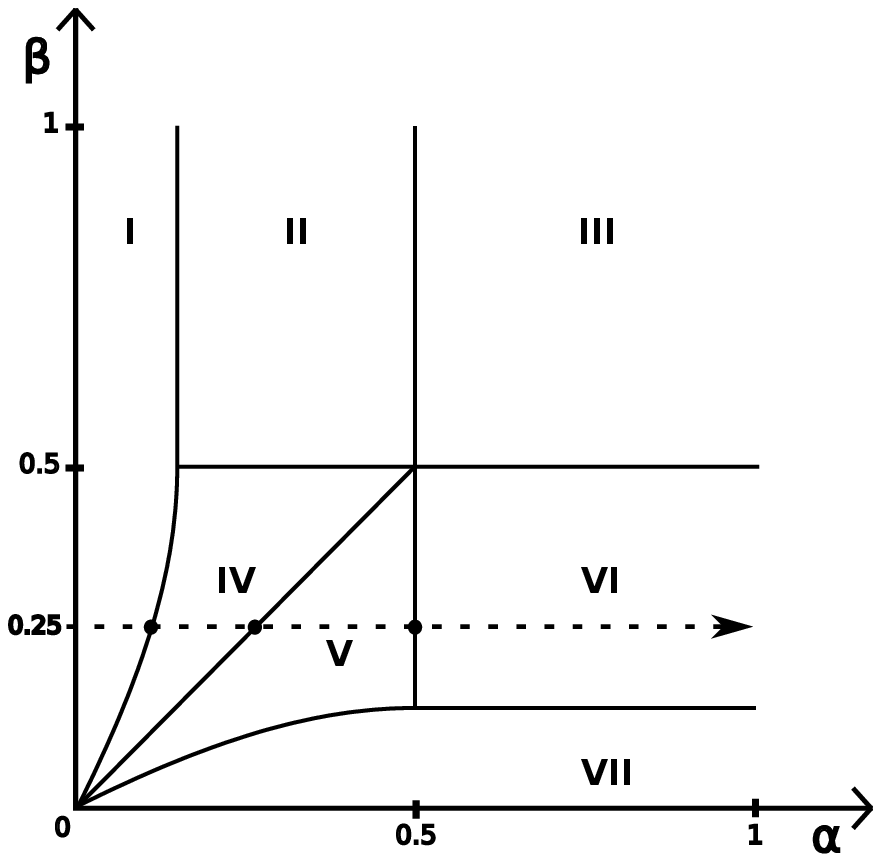}      
    &\includegraphics[width=0.3\textwidth]{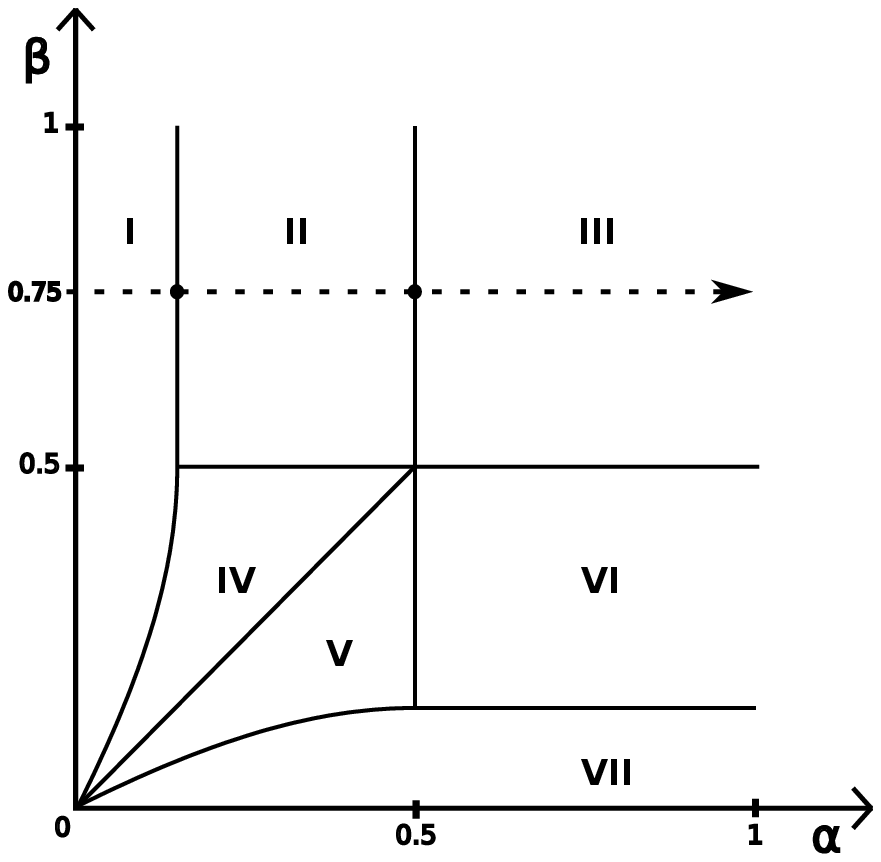}      
    \\
    \rotatebox{90}{\parbox{4cm}{\centerline{$J$}}}
    \hspace{-1cm}
    &  \includegraphics[width=0.38\textwidth]{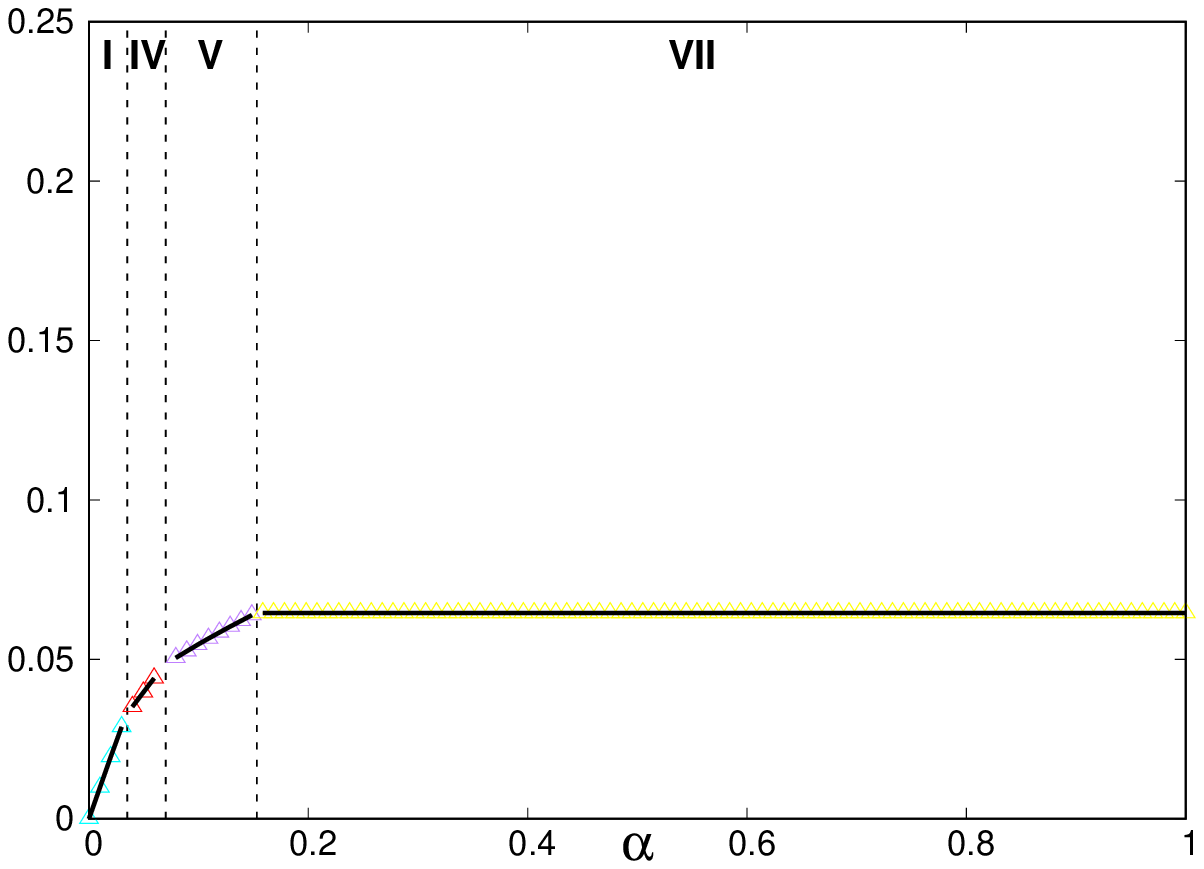}
    \hspace{-0.8cm}
    &  \includegraphics[width=0.38\textwidth]{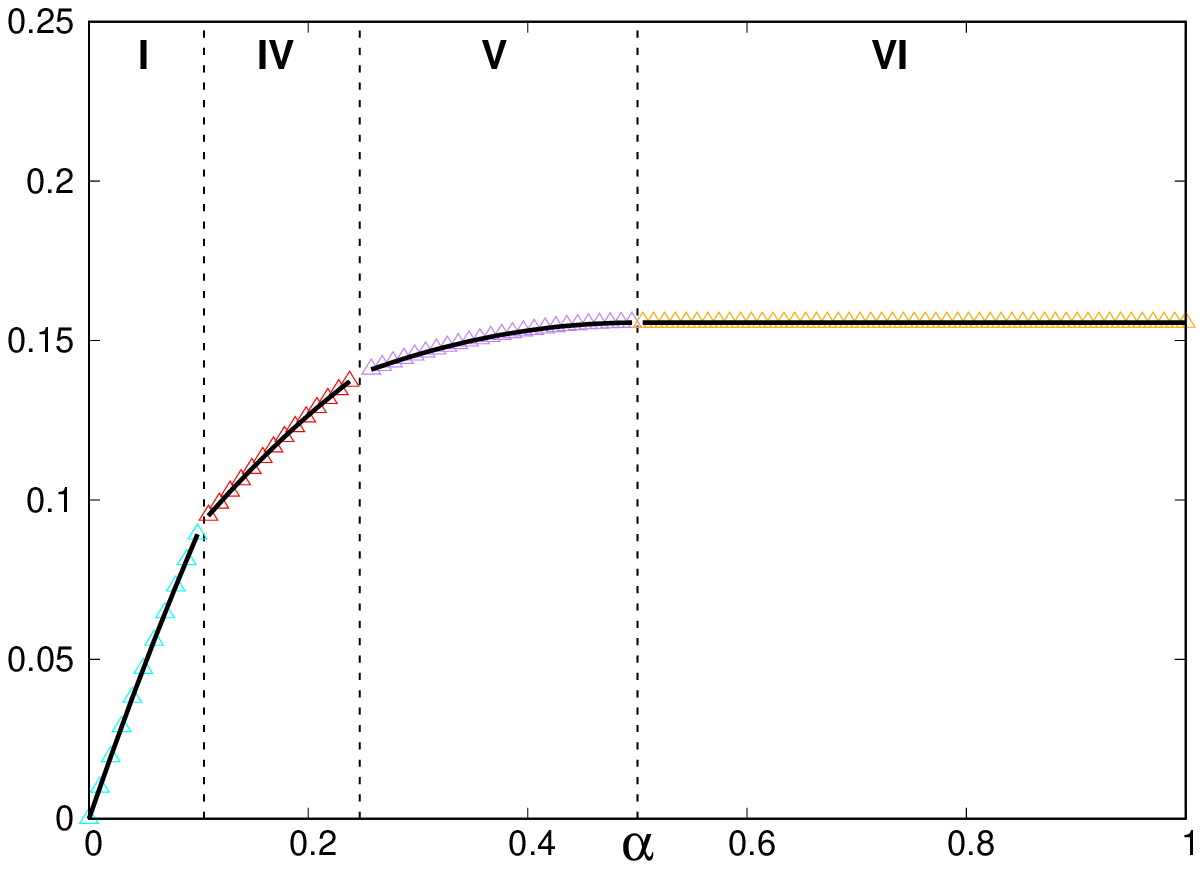}
    \hspace{-0.8cm}
    &  \includegraphics[width=0.38\textwidth]{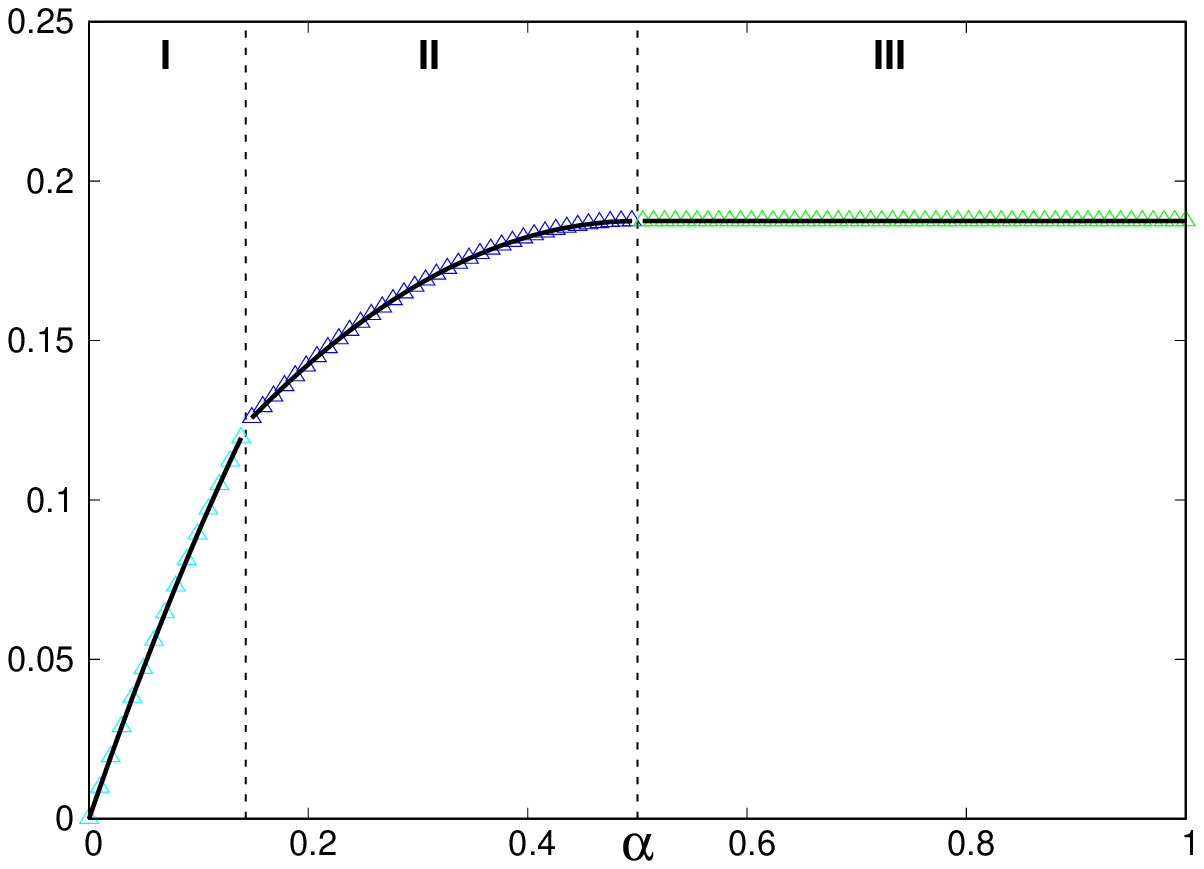}
    \\
    \rotatebox{90}{\parbox{4cm}{\centerline{$\rho_A, \rho_B $}}}
    \hspace{-1cm}
    & \includegraphics[width=0.38\textwidth]{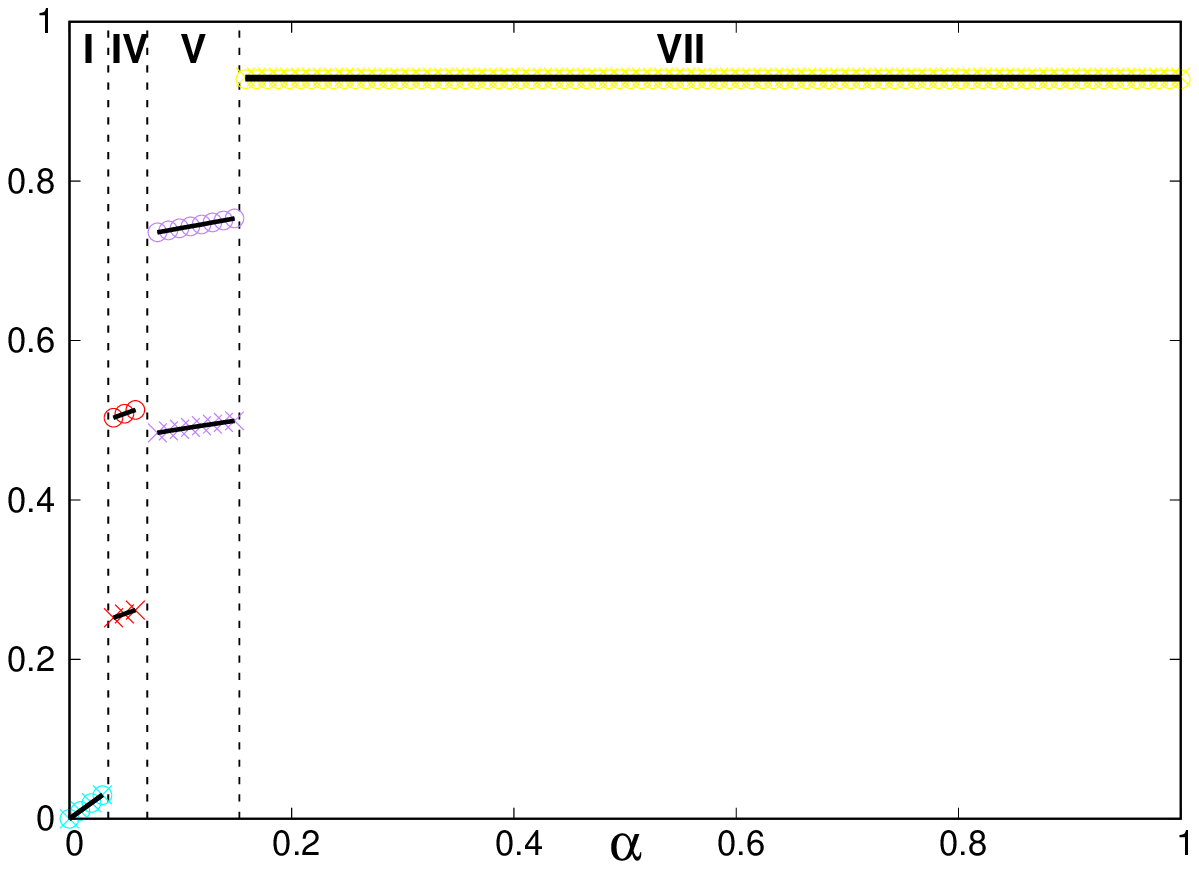}
    \hspace{-0.8cm}
    & \includegraphics[width=0.38\textwidth]{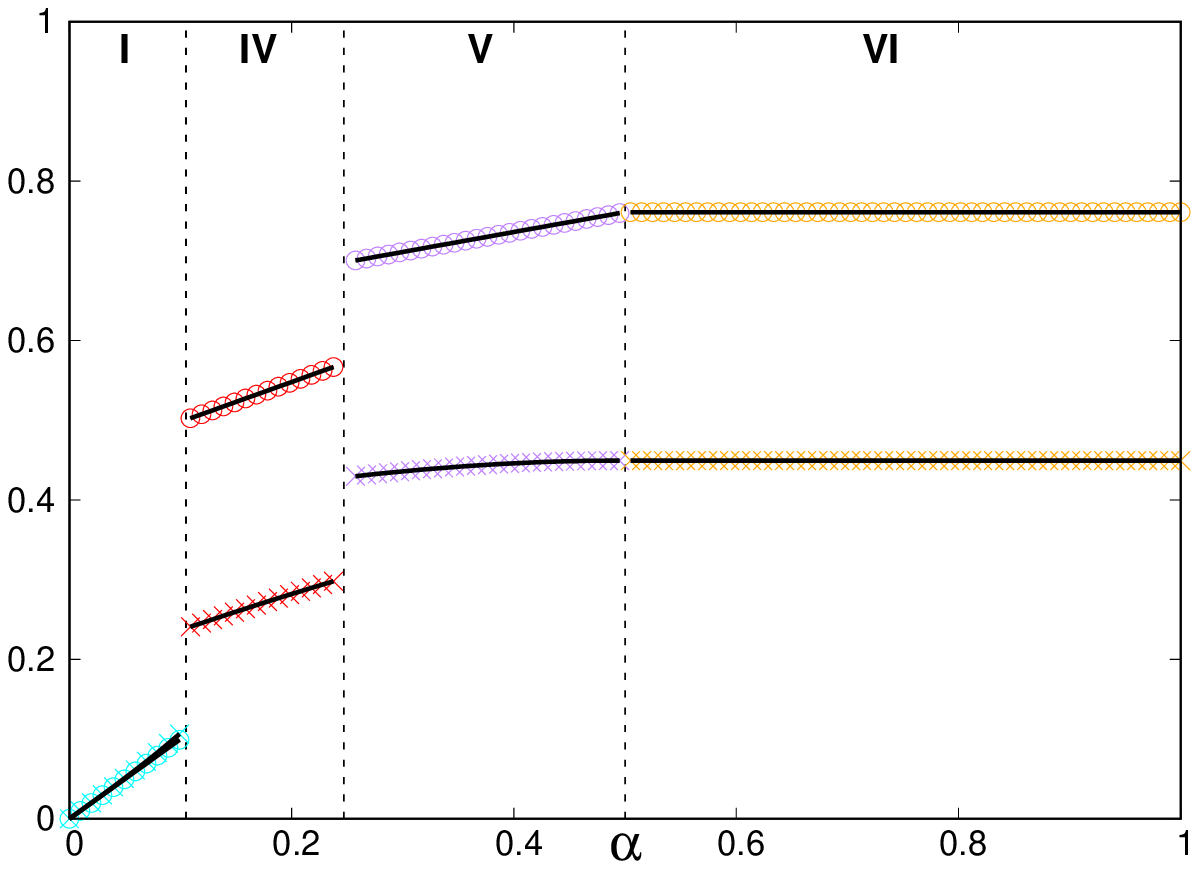}
    \hspace{-0.8cm}
    & \includegraphics[width=0.38\textwidth]{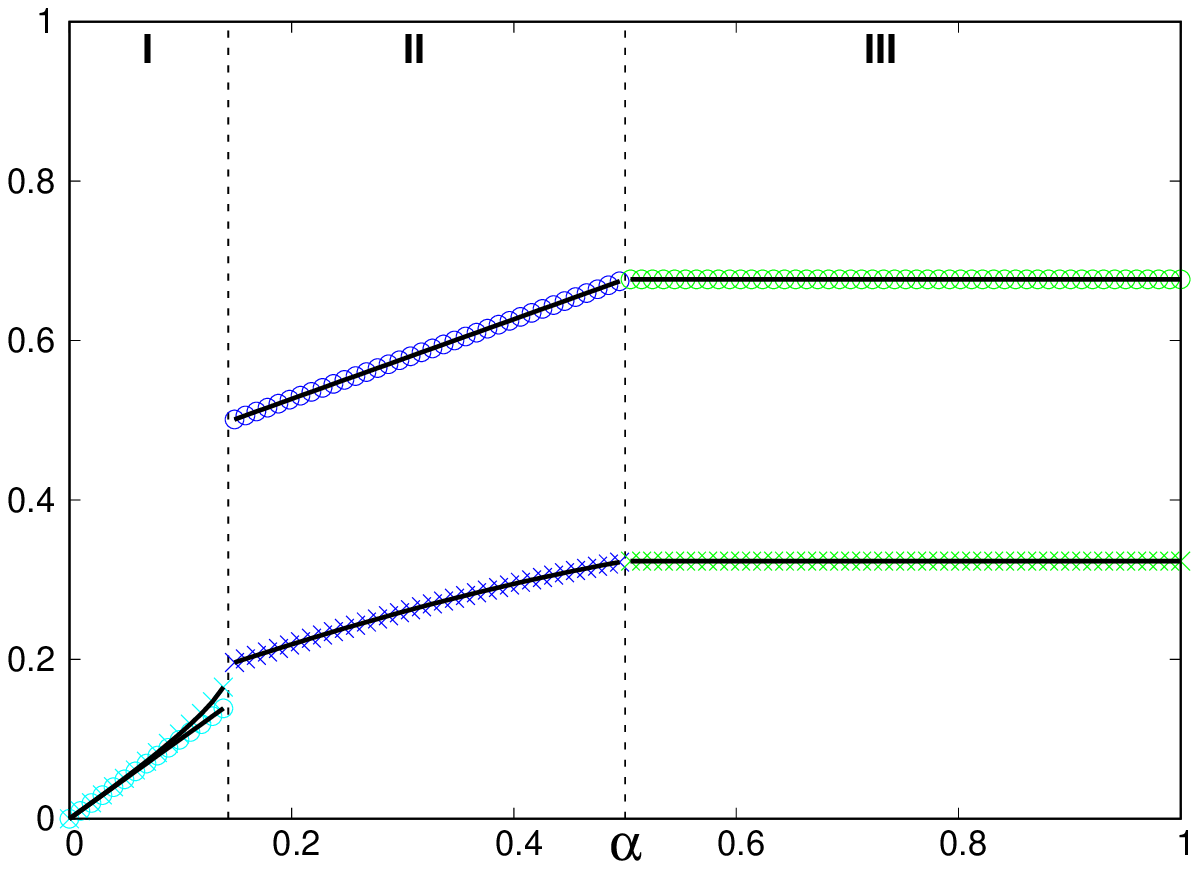}
  \end{tabular}
  \caption{
    (colour online)
    Current ($J$) and densities upstream ($\rho_A$) and downstream ($\rho_B$) from the full crossing, at fixed exit rate ($\beta$): $\beta=0.07$, $\beta=0.25$, and $\beta=0.75$. Simulations are plotted with coloured symbols, both for the current and the densities. Colours code for each transport phase, additional labels are as defined in Figure \ref{diagramme_superpose}.
    Empty circles represent the density upstream from the crossing ($\rho_A$) while crosses represent the density downstream from the crossing ($\rho_B$). The current $J$ is plotted with empty triangles. 
    Note that the spacing of the data points has been maintained close to the critical values of $\alpha$ (dashed lines). In all cases the current is a continuous function of the entrance rate, jumps in the density indicate discontinuous transitions.
}

  \label{beta_fixed}
\end{figure*}

We have therefore constructed such 'sweeps' in the entry rate $\alpha$, for several values of the exit rate $\beta$, chosen to cover the three possible scenarios. The results are shown in Figure \ref{beta_fixed}, where both the upstream and downstream density on the microtubule are represented, as well as the current, as a function of $\alpha$. The current $J=J_{A,B}(\alpha)$ is thus seen to be a monotonously increasing function of $\alpha$ which reflects the phase boundaries in the compound phase diagram. The densities $\rho_A$ and $\rho_B$ differ for most phases. Their values may or may not experience a 'jump' at critical values of the entrance rate, which illustrates that transitions may be discontinuous or continuous. The number of observed transitions depends  on the value of the exit rate, as is easily rationalised from the phase diagram in Figure \ref{beta_fixed}, according to the path followed in the $(\alpha,\beta)$ plane. Interestingly, transitions can be continuous in the downstream density $\rho_B$ but discontinuous in the upstream density $\rho_A$, as exemplified in the transition (I-II) for $\beta=0.75$.

\subsection{Interpreting the compound current-density relation}

One relevant way of envisaging the current of motors through a microtubule crossed by another one is to think of it as a coarse-grained transport path, with a current-density relation for this compound object. We will construct this now, based on the densities specified in Eqs. (\ref{eq:rhoA}) and (\ref{eq:rhoB}), as well as the corresponding mean-field currents determined from Eq. (\ref{current_density}).

\begin{figure*}
  \subfigure[]{
    \centering
    \includegraphics[width=0.8\textwidth]{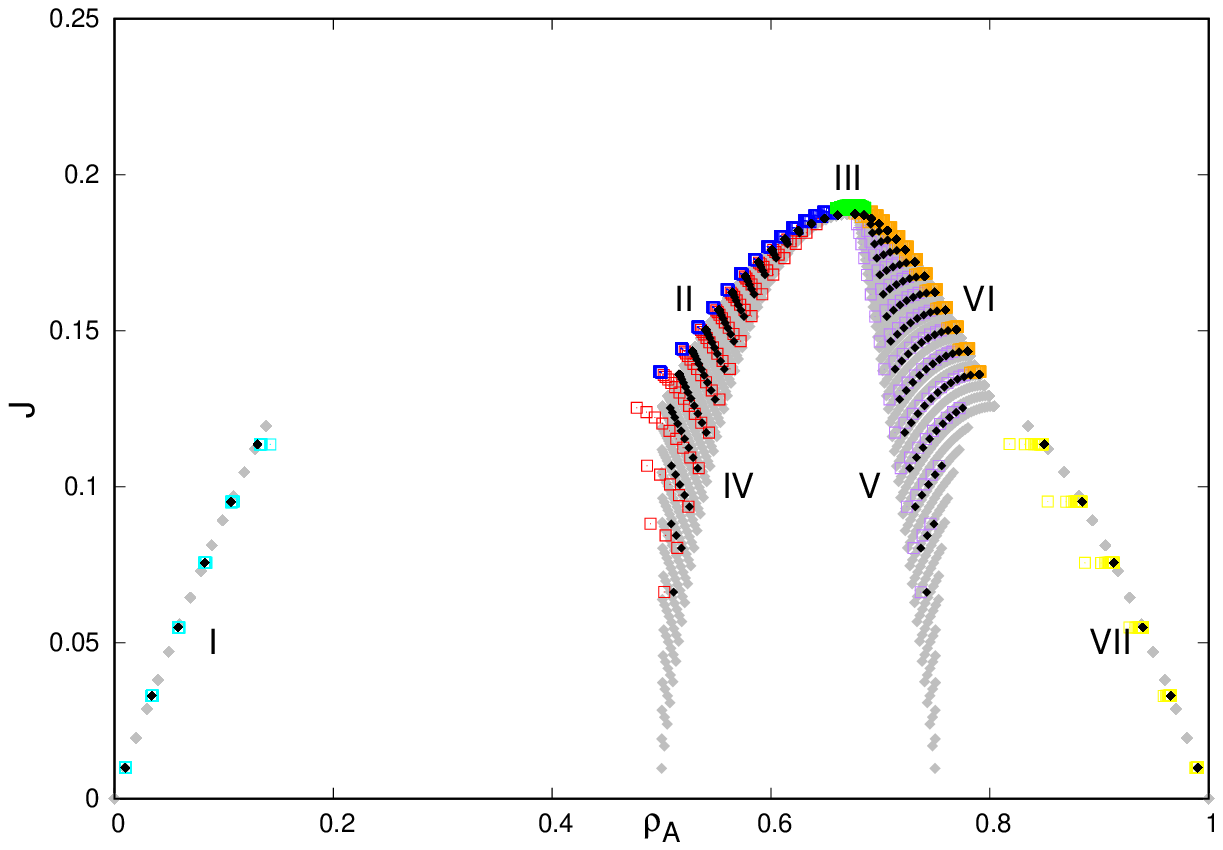}
    \label{current A}
  }
  \hspace{1.0cm}
  \subfigure[]{
    \centering
    \includegraphics[width=0.8\textwidth]{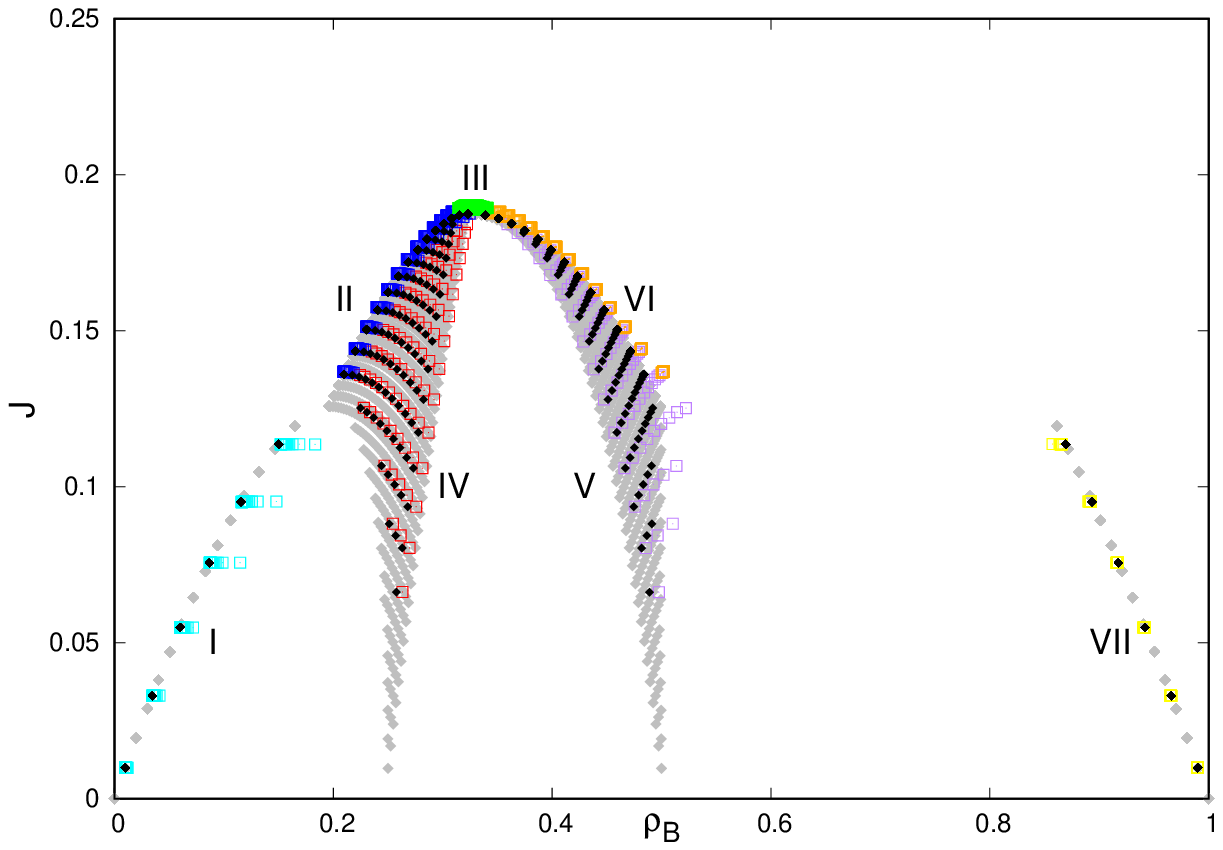}
    \label{current B}
  }
  \captionof{figure}{
    (colour online)
    Locally averaged current-density relations upstream ($J(\rho_A)$) and downstream ($J(\rho_B)$) the crossing point. Grey diamonds show a dense enumeration of mean-field predictions for pairs of boundary rates ($\alpha$,$\beta$) of the parameter plane. Open squares show simulation data for a selection of boundary rates pairs. Black diamonds show the analytic mean-field prediction corresponding to each selected ($\alpha$,$\beta$) pair. Open squares are colour-coded depending on the transport phase, which are additionally indexed by the capital roman labels as in Figure \ref{diagramme_superpose}: respectively for vertices V(1:2), V(2:1) and V(1:1) we have I (LD:2LD; 2LD:LD; LD:LD), cyan; II (LD:2LD; 2HD:MC; LD:LD), blue; III (MC:2LD; 2HD:MC; MC:MC), green; IV (LD:2LD; 2HD:HD; LD:LD), red; V (LD:2LD; 2HD:HD; HD:HD), purple; VI (MC:2LD; 2HD:HD; HD:HD), orange and VII (HD:2HD; 2HD:HD; HD:HD), yellow.
  }  
  \label{current A and B}      	
\end{figure*}

For each pair of rates $(\alpha,\beta$), the resulting compound current and densities are calculated from mean-field expressions, and represented in the two plots of Figure  \ref{current A and B}, one for the upstream and one for the downstream measurement zone, where each data point corresponds to one pair of values $(\alpha,\beta)$. In order to provide an overall view of the data, but also to allow for comparison with simulation data, we use three complementary symbols: (i) light gray diamonds show all mean-field data points from a rather dense enumeration of points in the $(\alpha,\beta)$ parameter plane; (ii) open squares show data as obtained from Monte Carlo simulation with  specific choices for the rates $(\alpha,\beta)$, selected throughout the parameter plane; (iii) black diamonds show the mean-field predictions for exactly those parameters used for simulations, in order to allow for direct comparison. Labels carrying roman numbers correspond to those identifying the phases in Figure \ref{diagramme_superpose}.

A direct conclusion can be drawn from comparing simulation data with the corresponding mean-field results. It becomes clear that the predictions, if not entirely accurate, are overall largely correct: simulation points (open squares) systematically fall close to, if not on top of, mean-field predictions (black diamonds).

A first point is that the average current can never reach the maximum value of 1/4, as it is impossible to simultaneously find all protofilaments in a MC phase.
But there are several other features of the current-density relation which deserve attention.
Each of them highlights a central observation: the current-density relation Eq. (\ref{current_density}) is qualitatively different from what would be expected from a {\it coarse grained} picture. Indeed, intuitively one might expect that a coarse-grained description would be able to describe transport through the  microtubule crossing in terms of a mapping onto a coarse-grained  V(2:2) branching point between two transport paths, the properties of which are well known \cite{embley2009}. However, this is {\it not} the case: the microscopic routing through the ensemble of transport paths confers the compound transport relation distinctive new properties.

We discuss some of these features in the following.

\paragraph{Density gaps.}
First, the current-density shows 'gaps', {\it i.e.} certain density ranges cannot be attained. It has been illustrated before that this is a feature of an open system, for which the motor density can adapt freely to the imposed entrance/exit rates \cite{embley2008, embley2009}. The existence of these gaps is directly due to the presence of branching in the microscopic transport pathways.

\paragraph{Degeneracy of the current-density relation.\label{sec:degeneracy}}
  
Second, the current-density relation is strikingly different from what is usually observed, including for branched systems: we are not dealing with a one-to-one relation, {\it i.e.} the current is not a well-defined function of the density. Indeed, the current-density relation traces out an entire surface, rather than a line. 

Instead, simulation points for the locally averaged relations between currents ($J_A$ and $J_B$) plotted as a function of the corresponding densities ($\rho_A$ and $\rho_B$) are spread out over surfaces.
This amounts to an infinite degeneracy for certain phases, as a  \textit{continuum} of values of the current corresponds to a single value of the density, or vice versa.

To see how this is due to an ambivalence in the boundary rates we focus on phase $V$ (LD:2LD; 2HD:HD; HD:HD) as an example. In this phase, the (averaged) current in the compound tubule can be read from Table \ref{table:vertices} as
\begin{equation}
  \label{eq:compound:JA:example} 
  \begin{split}
    J & =\frac{1}{4}[
      J_{A_,V(1{:}2)} + 2J_{A,V(2{:}1)} + J_{A,V(1{:}1)}]\\
    & =\frac{1}{4}[\alpha(1-\alpha)+2\beta(1-\beta)]
    \ .
  \end{split}
\end{equation}
Similarly, the density on the upstream compound tubule can be read off as
\begin{equation}
  \label{eq:compound:rhoA:example}
  \begin{split}
    \rho_A
    & =\frac{1}{4}[ \rho_{A,V(1{:}2)} + 2\rho_{A,V(2{:}1)} + \rho_{A,V(1{:}1)} ]\\
    & =\frac{1}{4}[ 2+\sqrt{1-2\beta(1-\beta)} + \alpha -\beta ]
    \ .
  \end{split}
\end{equation}
The key here is that the rates $\alpha$ and $\beta$ may be chosen independently, within the intervals compatible with this phase. Say these parameters are chosen to produce a (compound) current $J=J_A$ at some (compound) density $\rho_A$. Changing either $\alpha$ or $\beta$ will affect these values. However, it is now possible to match the changes in these rates such that the density $\rho_A$ does not change, although the current still does: it suffices to accompany a variation in $\alpha$ by the required change in $\beta$, chosen to maintain the equality expressed by Eq. (\ref{eq:compound:rhoA:example}). However, this will typically modify the compound current $J=J_A$  (Eq. (\ref{eq:compound:JA:example})). Therefore there is a {\it continuum} of current values which may be produced at a given density, by choosing the entrance and exit rates as required.
Similarly, an infinity of densities may be selected which all produce the same current.

\paragraph{Questions of particle-hole symmetry.}
Third, both the upstream and the downstream current-density relations do not show a particle-hole symmetry, {\it i.e.} the interval of currents which can be achieved for $\rho_A$, for example, differs from the one for $1-\rho_A$. They are, however, symmetric one with respect to the other when the mapping $\rho_A \leftrightarrow 1-\rho_B$ is made. This is a direct consequence of the symmetry in the compound phase diagram, already discussed in section \ref{sec:symmetry}.

\paragraph{Location of the zone of measurement.}

Finally, an important point follows directly from the simple observation that the upstream and downstream relations are not identical: the location of the zone of measurement matters.
This is also an entirely novel observation, which is due to the compound nature of the microtubules, as well as that of the crossing. Specifically, the current-density relation which one would observe by measuring current and density is different according to whether an upstream section or a downstream section of the microtubules are probed.

\section{Discussion and conclusions}

Essential features of cell structure and functionality rely on the cytoskeleton and on cytoskeletal transport. This is a collective process,  in a crowded environment, taking place on the scaffolding of interlinked filaments which guide molecular motors.
This transport process is crucial, yet \textit{in vivo} and \textit{in vitro} experiments are still very challenging. This is mainly due to the fact that we are dealing with a large scale process, taking place in the complex environment which is the cytoplasm, as well as its dynamical nature. 

\paragraph{Summary and discussion}

In this work we have addressed the question how the structure of crossings between microtubule filaments, where motor proteins may switch from one filament to another, affect the collective transport. We have done so by considering the crossing as a compound object, which we have decomposed into individual 'paths' along which molecular motors advance stochastically. Rules for switching between such paths have been formulated to represent a schematic microtubule crossing, but can easily be adapted to other scenarios. We have then implemented the TASEP model for transport, a simple but paradigmatic model, in order to highlight how the internal structure of the compound crossing strongly affects the features of transport.

We have constructed the full phase diagram of the average density of the crossing filaments by a mean-field method, based on well-established results for TASEP transport on branched structures. It features distinctive phases, according to how the entry and exit rates control individual transport paths. Both continuous and discontinuous transitions in the average density are present. Direct numerical simulations of the transport process corroborate our findings. 

A central result is the corresponding current-density relation. This relation is key in traffic sciences for characterising the capacity of a transport system, as it quantifies its ability to convey matter efficiently \cite{schadschneider2000}. Here we found that the fundamental relation shows several new features. These  set the compound crossing apart from a single branching point between transport paths, as they have been considered heretofore. Most strikingly, there no longer is an unequivocal relation between current and density. Instead, in certain density regimes, an entire range of currents can be produced for the same density by varying the entry and exit rates. This is qualitatively different from what is known for TASEP transport, even when branching points are present. The phenomenon is directly linked to the fact that density and current are convoluted over all transport paths at a given point along the filament.  Note that this reflects the experimental situation, as resolving collective transport along each of the individual paths appears currently out of reach. Beyond the specific TASEP model used here, we expect this to be a generic property which should hold for any low dimensional transport process with compound crossings between filaments.

Importantly, it follows directly from this result, and also from the absence of particle-hole symmetry in the current-density relation, that it is impossible to represent the compound crossing as a coarse-grained 4-fold crossing between the two filaments. Rather, the internal structure of the transport paths through the crossing confers it a complexity which largely exceeds that of a simple branching point, and which is also reflected in the current-density fundamental relation.

Another important point is the observation that the current-density relation differs according to whether measurements are made upstream or downstream from the crossing. This too is novel, and differs from what is seen for a simple branching point. This feature is of particular interest, as it directly impacts measurements, which therefore need to consider carefully at which point the flux of motors is measured as a function of the motor average density.

\paragraph{Towards realistic microtubule crossings}

\begin{figure*}
  \hspace{1.0cm}
  \subfigure[]{
    \centering
    \includegraphics[width=0.5\textwidth]{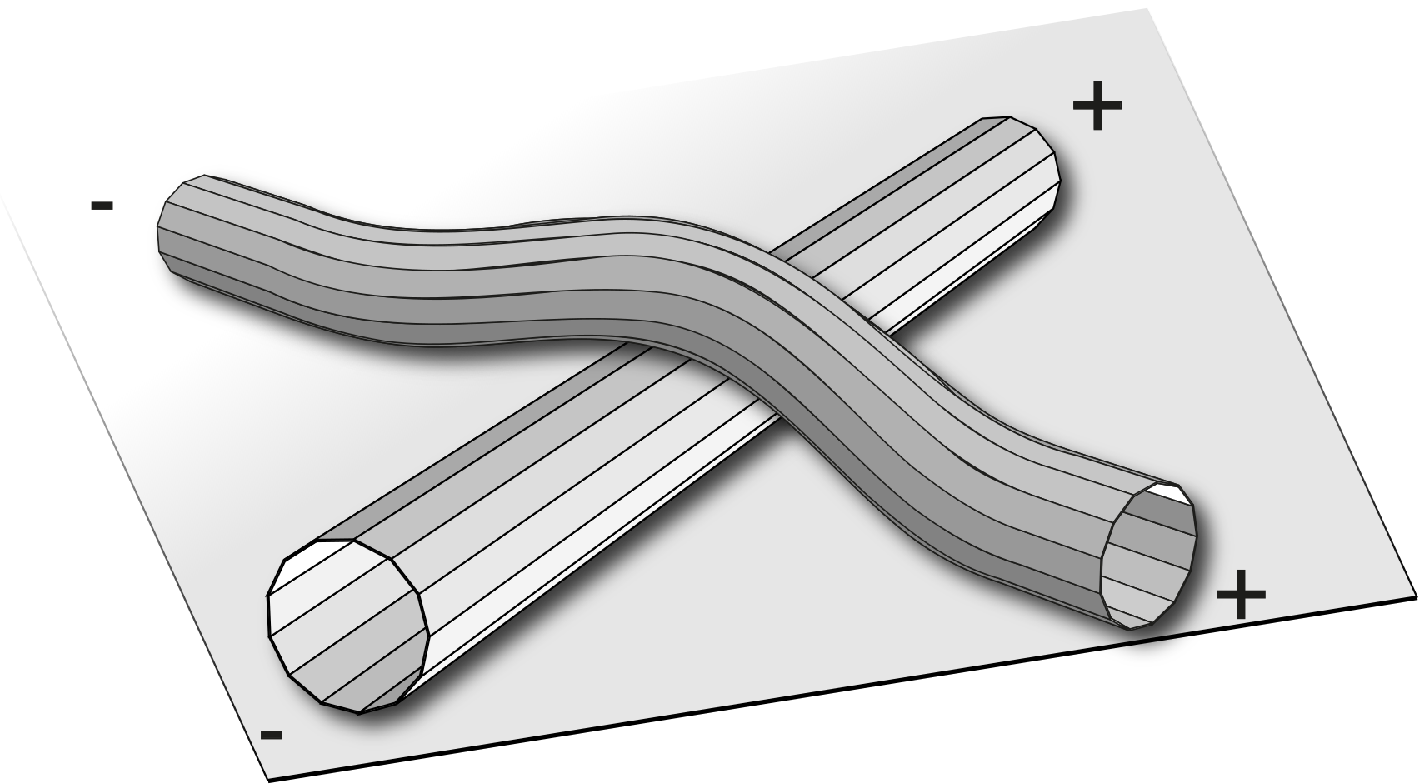}
    \label{fullcrossing:sketch}
  }
  \subfigure[]{
    \centering
    \includegraphics[width=0.4\textwidth]{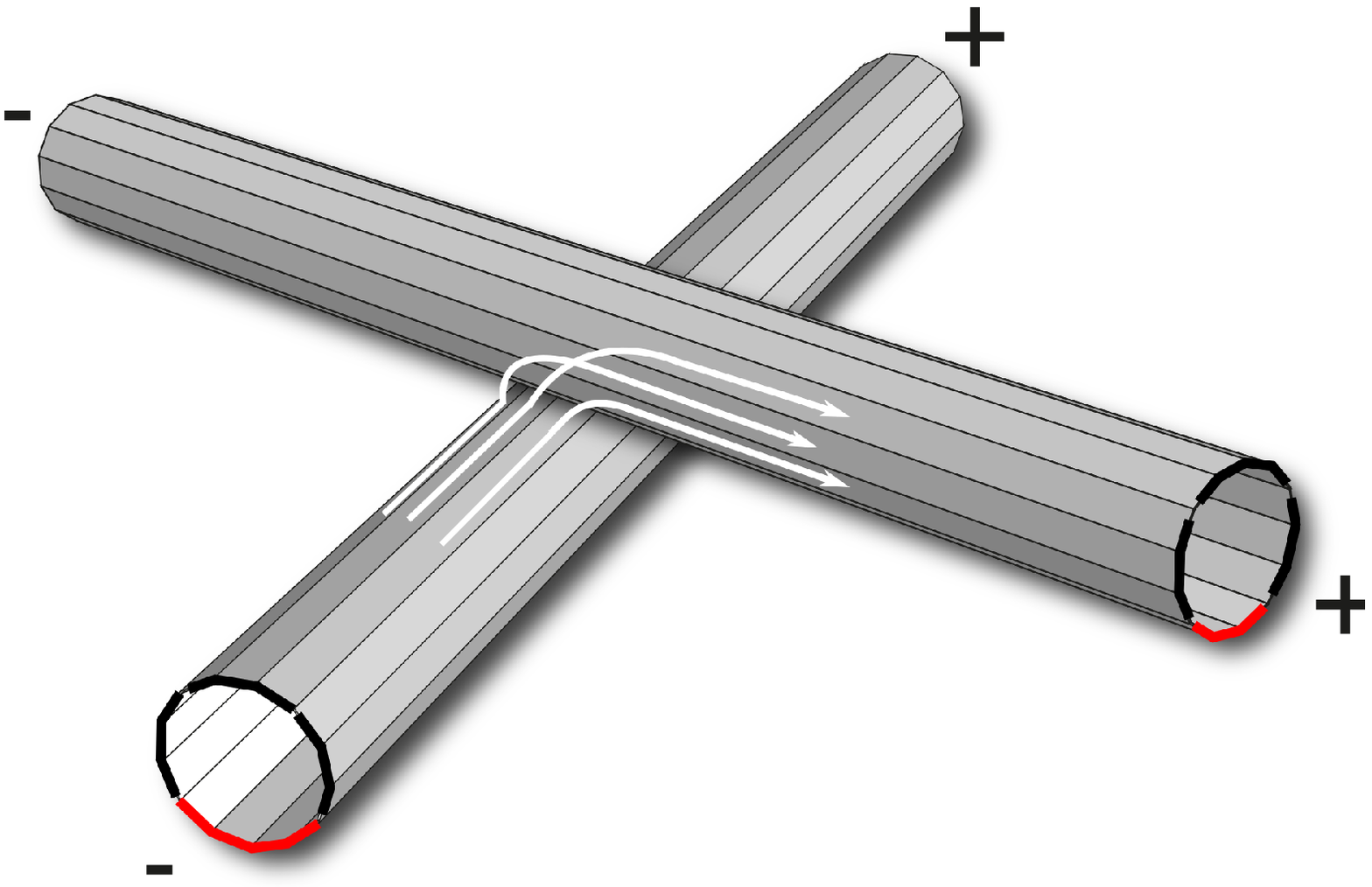}
    \label{fullcrossing:paths}
  }
  \captionof{figure}{
    (a) Schematic representation of a crossing of two full microtubules with 13 protofilaments, adhering to a glass surface. The scenario is inspired by the situation found in recent experiments \cite{deeb2019}.
    (b) A simple scenario for decomposing transport paths, for clarity shown on non-curved microtubules. Protofilaments close to the glass plate (shown in red)  are considered inaccessible, and for simplicity motors switching onto the alternative microtubule are taken to do so onto specific protofilaments. 
  }
\end{figure*}

In a larger context, the impact of our findings goes far beyond the, deliberately schematic representation, consisting of a 'microtubule' with 4 protofilaments, which we have employed to expose the new features as clearly as possible.
%
As a {\it proof of concept}, let us illustrate how information may be obtained 
on a biologically more realistic crossing of two full microtubules with 13 protofilaments.
To do so we elaborate one simple scenario, which is again inspired by the situation in the experiments by Deeb et al. \cite{deeb2019}, where microtubules are synthesised adhering to glass plates. We reason in terms of categories of protofilaments, according to how their transport is affected by the other microtubule: it may be blocked due to steric hindrance, receive or donate motor proteins, or remain entirely unaffected by the presence of the other microtubule.
These are essentially what made us retain four faces in our toy model.
Here we furthermore consider that those protofilaments adhering to the glass surface are entirely unavailable for transport, and we focus on the top microtubule as an example. For the sake of simplicity we assume that there are three protofilaments of each category, as illustrated in Fig \ref{fullcrossing:paths}. The thirteenth protofilament is attributed to the class of protofilaments along which no motors are received or gained.

It is important to keep in mind that this is one scenario, which we considered simplest when two full microtubules with 13 protofilaments are retained. Other scenarios may emerge as further details of the microscopic processes are discovered. Our aim here is simply to illustrate that a model with such complexity may be addressed, and to show that it leads to interesting predictions in terms of the routing process for cargos.

\begin{figure*}
 \subfigure[]{
    \centering
    \includegraphics[width=0.3\textwidth,angle=-90]{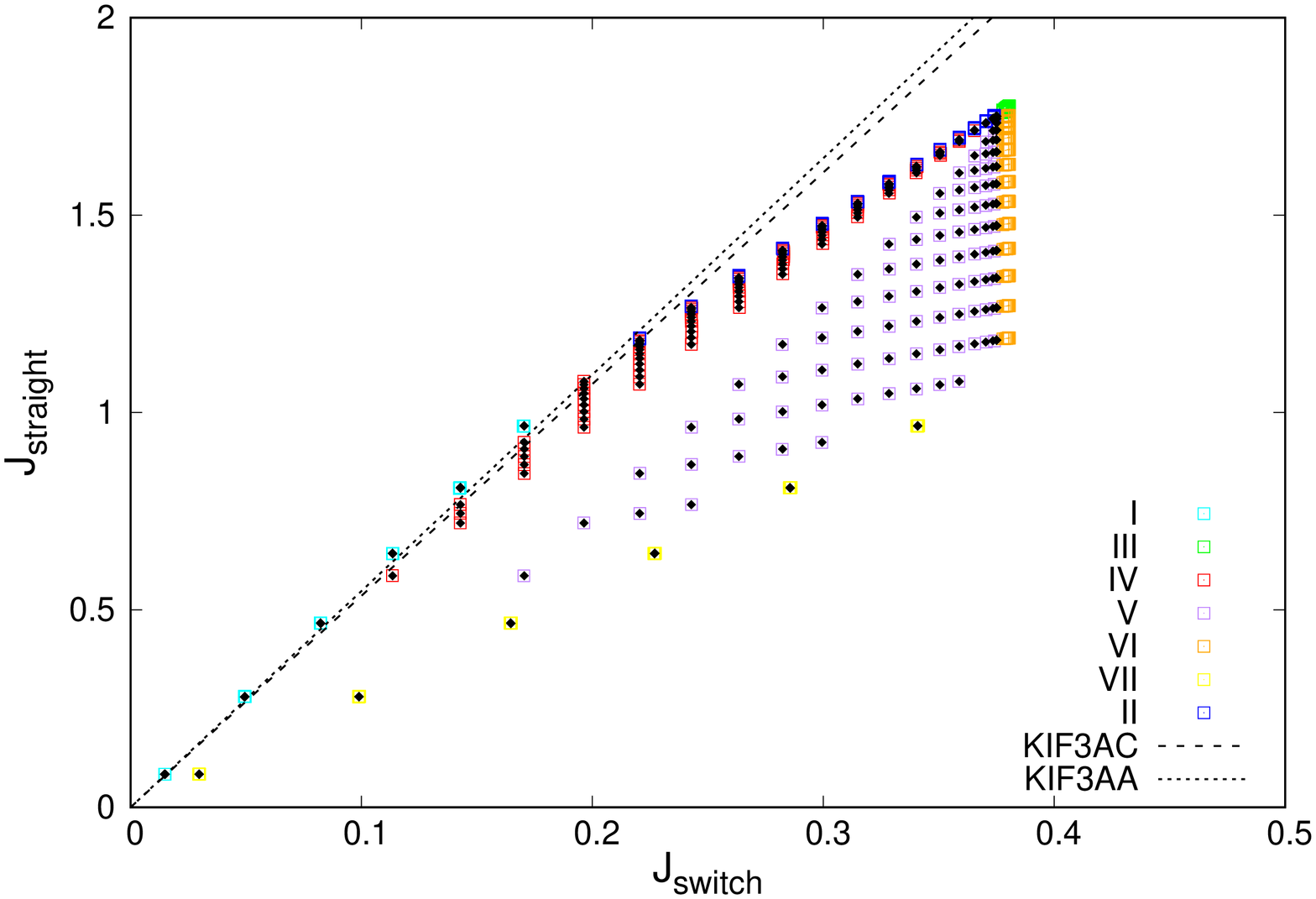}
    \label{fig:fullcrossing:plot}
  }
 \subfigure[]{
    \centering
    \includegraphics[width=0.3\textwidth,angle=-90]{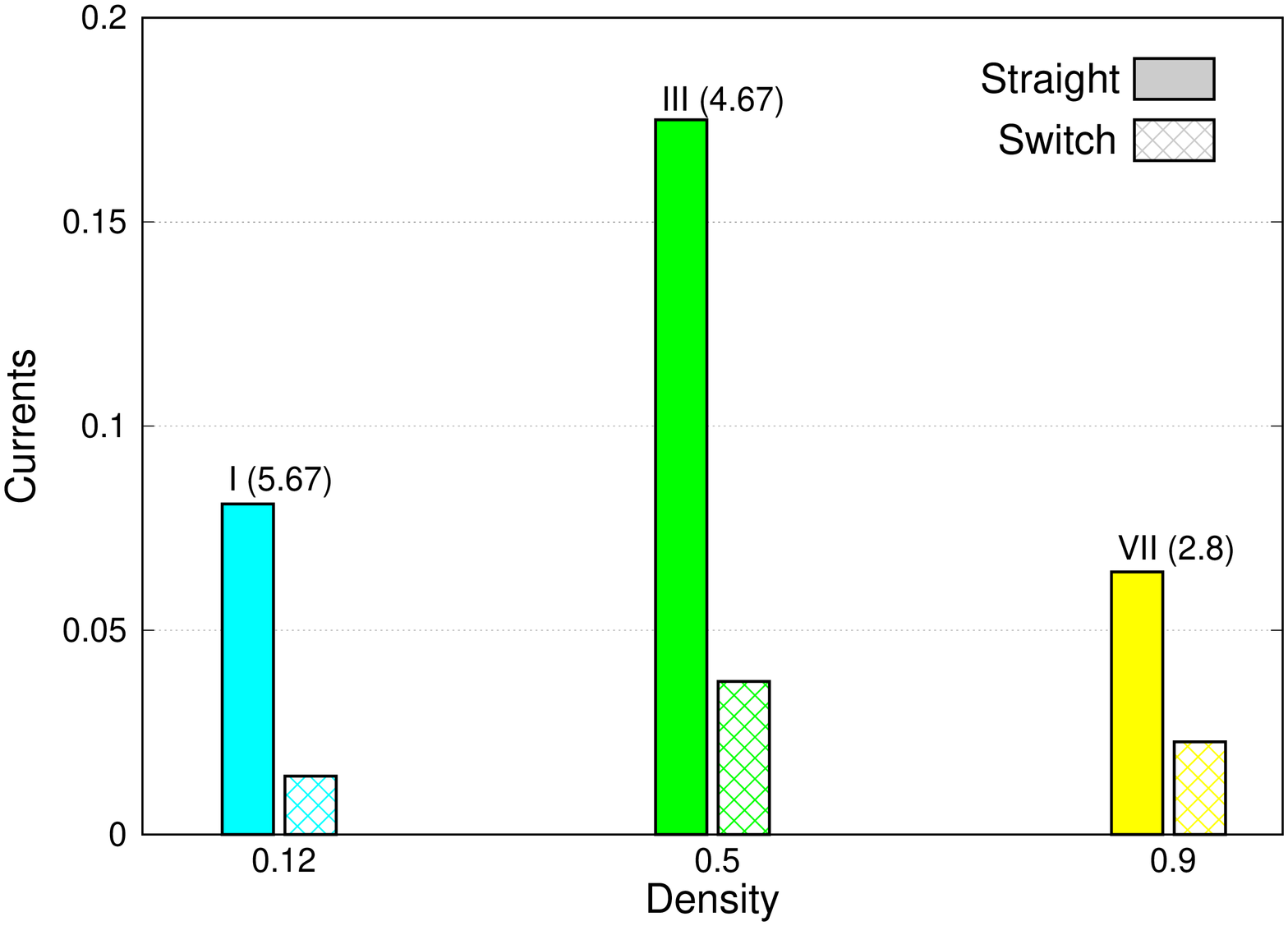}
    \label{fig:fullcrossing_histogram}
  }
 \captionof{figure}{
   (colour online)
   Results for a full crossing of microtubules with 13 protofilaments, based on the scenario sketched in Fig. \ref{fullcrossing:paths} and described in the text. The representation of results is inspired by the experimental work in \cite{deeb2019}, and focusses on the probability of 'straight' and 'switch' events at the crossing.
   (a) Each data point in the 'straight' vs. 'switch' plane corresponds to a particular choice of entrance/exit rates ($\alpha,\beta$), and the corresponding phases are identified according to the same symbols and colour code as in Figs. \ref{current A and B}. The straight line corresponds to experimental data from \cite{deeb2019} on Kif3AA.
   (b) Histograms representing the importance of 'straight' and 'switch' events, for different densities. Also indicated is the phase to which transport corresponds, and the ratio of straight vs. switch events. The importance of these events is seen to vary, indicating that re-routing is non-trivially affected by collective transport effects.
  \label{fig:fullcrossing:results}
} 
\end{figure*}

We therefore analyse the individual transport paths as they emerge from Fig.  \ref{fullcrossing:paths}. For the top microtubule, for example, the topmost protofilaments are unaffected, i.e. do not receive or donate motors: these are the ones which, in the light of our previous analysis, would have behaved as V(1:1). But we also see three paths corresponding to V(2:1) which receive motors from the other protofilaments at the crossing, and three V(1:2) which donate motor molecules, as well as three protofilaments which adhere to the surface and are therefore barred from transport. 

Currents and densities are then calculated by straightforwardly generalising Eq. (\ref{eq:compound:JA:example}) and (\ref{eq:compound:rhoA:example}), by adapting coefficients to account for all 13 protofilaments. Results are presented in Fig. \ref{fig:fullcrossing:results}, in a way to make contact with the experimental results in \cite{deeb2019}. This particularly focusses on the importance of 'straight' vs. 'switch' events, according to whether a motor protein continues to proceed along the same microtubule or switches over to the other microtubule at the crossing (recall that events where motors detach at the crossing were found to be relatively rare in the experiments and thus are not considered here).

Fig. \ref{fig:fullcrossing:plot} represents separately the currents of motors as they switch to the alternative microtubule ($J_{switch}$) and as they persevere on the same microtubule ($J_{straight}$). These currents correspond to the number of motors moving through the crossing per unit time, with or without switching, respectively. The depicted currents are therefore proportional to the number of corresponding events, and displaying them in the parameter plane of 'straight' vs. 'switch' characterises their relative importance, as exploited in \cite{deeb2019}. The colour code and symbols, identical to those in  Figs. \ref{current A} and \ref{current B}, furthermore indicate the underlying phases at which transport is found to operate for each point (the phases established in the phase diagram Fig. \ref{diagramme_superpose} are, by construction, independent of the multiplicity of each type of vertex).
The data points are produced by an enumeration of combinations of $(\alpha,\beta)$ rates, and therefore cover the entire density range.
Interestingly, the low density data (labelled as phase I) follow a straight line, which furthermore corresponds to the correlation determined in \cite{deeb2019}, where straight  passing events have been found to be roughly 5 times more likely than switching events. Although this does not prove that the hypotheses underlying our scenario are correct or complete, it does mean that our model clearly meets the experimental result for very low motor concentrations, as any valid model should do.

Another point is illustrated in Fig. \ref{fig:fullcrossing_histogram}, where the importance of 'straight' and 'switch' events is characterised, opposing several motor densities.
The main point is that the  importance of these events varies depending on the density. Recall that molecular motors were assumed to select any path which is open to them with equal probability. However, the occupancy differs between the different types of protofilaments, and can thus render either 'straight' or 'switch' events more likely, and this effect depends on density. The process of routing cargos through a crossing is therefore seen to be sensitive to density, in a non-trivial way. For motor densities as they are expected to arise in crowded cellular environments, the routing process is therefore impacted significantly by collective transport effects:
the ratio of straight vs. switch events varies by about a factor of 2 for the densities represented in Fig. \ref{fig:fullcrossing_histogram}.

\paragraph{Outlook} 
The overall context holds many further perspectives.
Clearly, the model presented here has omitted many aspects of the complexity of the processes which govern how motor proteins behave at microtubule crossings. Many improvements can be envisaged, and several such refinements would be of particular interest in the light of modelling actual cytoskeletal transport. Indeed, it should be straightforward to incorporate the absorption/desorption of molecular motors (using the TASEP-LK model, as has been done in \cite{neri2013}). Back-stepping of motors can be addressed in the same framework (also covered in \cite{neri2013b}). More complex microscopic behaviour at the crossing would require differentiating transition rates for switching between filaments, such as a bias (as in \cite{embley2009}) or particular stepping rates at the crossing itself (as in \cite{raguin2013}). Further aspects which would be interesting to study comprise bi-directional transport \cite{muller2008} and the impact of motors switching between protofilaments even far away from the crossing \cite{reichenbach2007, melbinger2011}, or several motors pulling a cargo \cite{erickson2011}, all of which may arise in biological systems.

On a more exploratory note, it may be worth pointing out that crossings between cytoskeletal filaments are in fact dynamic structures. Indeed, various protein complexes intervene in many ways in the sophisticated assembly, and affect or initiate processes such as filamental growth, branching and disassembly, as well as modify transport along the filaments themselves \cite{howard2003, conde2009, okada2015, goodson2018}. As we have seen that crossings affect the dynamics of the collective transport process in non-trivial manners, one may thus wonder to which extent regulating these protein complexes could also be a means of dynamically programming this formidable transport machinery, in order to implement the need of routing and re-routing biological cargos within the cell.

\section*{Acknowledgements}
We acknowledge financial support by the CNRS and by the Scientific Council of the University of Montpellier. The current position of A.R. is funded by the Federal Ministry of Education and Research of Germany in the framework of CornWall (project number 031B0193A).

\end{multicols}

\clearpage
\appendix

\section{Appendix on 3-fold vertices} \label{Appendix on 3-fold vertices}

In this section we review a generic method for constructing the mean-field phase diagram of TASEP segments interconnected at a branching point, based on the approach of effective rates \cite{embley2008, embley2009}. In order to do so, we denote the density on the branching point $\tilde\rho$ and then consider each TASEP segment individually. The idea is that the density $\tilde\rho$ sets effective entrance/exit rates for the joining segments, so that we can then exploit the phase diagram of a single TASEP segment (see section \ref{Collective transport in TASEP}). For the branched system, we must list all possible phase combinations, as each segment may potentially be found in either a LD, HD or MC phase.  Among these phase combinations, only those are to be retained which are compatible with current conservation through the crossing, for ranges of $\alpha$ and $\beta$ which are not mutually exclusive. 
  
For the purpose of illustration we focus on the 3-fold vertex V(1:2) and we detail the calculation of the phase LD:2LD. We hence decompose this branched structure comprising a 3-fold branching point into three individual TASEP segments. We aim, for example, to achieve that segment A (upstream from the crossing) as well as both segments B$_{1,2}$ (downstream from the crossing) be in the LD state. 
  
In terms of the density at the branching point $\tilde{\rho}$ (see Figure \ref{rates_V12}),  the effective rates to be considered are \cite{embley2009}
\begin{equation}
\left\{
    \begin{array}{ll}
        \beta_{eff}=1-\tilde{\rho}\\
        \alpha_{eff}=\frac{\tilde{\rho}}{2}\ .
    \end{array}
    \label{eq_rates}
  \right. 
\end{equation}
In essence, a particle can exit from the upstream segment A into the branching point requiring only to find that site unoccupied (probability $1{-}\tilde\rho$), which sets the rate $\beta_{eff}$. A particle on the branching point (present with probability $\tilde\rho$) can be supplied to either of the downstream segments B (probability $1/2$), which sets the rate $\alpha_{eff}$.
%
Based on these effective rates at the branching point, as well as the boundary rates $\alpha$ and $\beta$, we can now characterise each segment through its entry and exit rates, just as for a single TASEP segment. Therefore, the conditions required for setting a LD phase both on the segment A and the segments B$_{1,2}$ can now be read off the standard phase diagram of a single TASEP segment (see Figure \ref{parabole}).
\\

We find that, since segment A is required to be in the LD phase, its entry rate $\alpha$ must obey
\begin{equation}
\left	\{
    \begin{array}{ll}
\alpha < \beta_{eff}\\
\alpha < 0.5\
\end{array}
\right.
\xLeftrightarrow[]{\text{Eq. (\ref{eq_rates})}}
\left \{
	\begin{array}{ll}
	\alpha < 1-\tilde{\rho}\\
	\alpha < 0.5\ .
	\end{array}
	\label{effective rates segment A}
\right. 
\end{equation}
In addition, the density in a low density phase being equal to the entry rate, we thus have $\rho_{A}=\alpha$ and $J_A=\alpha(1-\alpha)$ (see section \ref{Collective transport in TASEP}).
\\

Similarly, for the downstream segments B$_1$ and B$_2$ to be in a LD phase, the conditions for the effective entrance and exit rates are
\begin{equation}
\left	\{
\begin{array}{ll}
\alpha_{eff} < \beta\\
\alpha_{eff} < 0.5\
\end{array}
\right.
\xLeftrightarrow[]{\text{Eq. (\ref{eq_rates})}}
\left \{
\begin{array}{ll}
\tilde{\rho} < 2\beta\\
\tilde{\rho} < 1\ ,
\end{array}
\label{effective rates segment B}
\right.
\end{equation}
Like for segment A, this implies that $\rho_{B}=\alpha_{eff} = \frac{\tilde{\rho}}{2}$ and $J_B=\frac{\tilde{\rho}}{2}(1-\frac{\tilde{\rho}}{2})$ (see section \ref{Collective transport in TASEP}).
\\

We can now deduce the remaining unknown of the system, the density at the branching point $\tilde{\rho}$, by imposing current conservation at the branching point
\begin{equation}
  J_A=2J_B \Leftrightarrow \tilde{\rho}=1-\sqrt{1-2\alpha(1-\alpha)}
  \ ,
\label{density at the junction}
\end{equation} 
where the latter condition is always true, since the negative branch must be chosen to respect $\tilde{\rho}<1$, and always offers a real solution for $\alpha<0.5$.
\\

Finally, we deduce the domain of existence of the phase LD:2LD in the parameter plane ($\alpha$,$\beta$) by combining all conditions on the rates $\alpha$ and $\beta$, as they are specified by Eqs. (\ref{effective rates segment A}) and (\ref{effective rates segment B}).  Expressing this in terms of the density $\tilde{\rho}$ at the crossing point via Eq. (\ref{density at the junction}), the condition for the phase LD:2LD requires rates $\alpha$ and $\beta$ such that
\begin{equation}
\left\{
\begin{array}{ll}
\alpha < 0.5\\
\beta > \frac{1-\sqrt{1-2\alpha(1-\alpha)}}{2}\ .
\end{array}
\right.
\label{eff_rates} 
\end{equation}

The reasoning is entirely analogous for all other phases.

\section{Mock-up\label{app:mockup}}

\begin{Figure}
	\centering	
	\includegraphics[width=\textwidth]{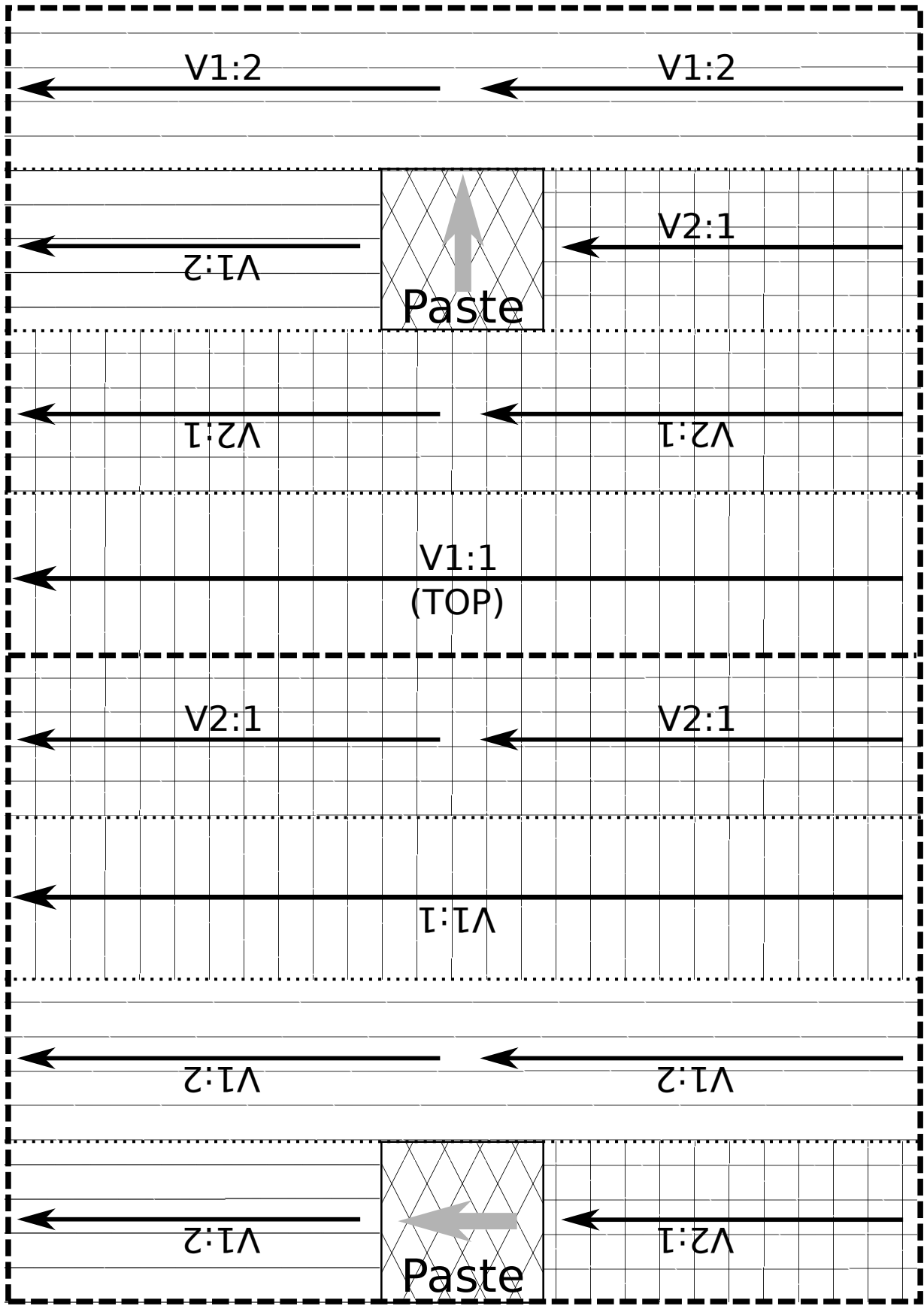}      
	\captionof{figure}{Mock-up of the 3-dimensional crossing of parallelepiped-like microtubules. To build it, simply cut along the dashed lines, fold along the dotted lines, and paste the two parallelepipeds such that the two grey arrows match. By orienting the face labelled with "(TOP)" towards the top, the unwrapping of the paths along the microtubules into vertices, following the Figure \ref{microtubules crossing}, becomes apparent immediately.}        
	\label{mockup}  
\end{Figure}


\begin{thebibliography}{99}

\bibitem{hirokawa2010} Hirokawa, N., Niwa, S., Tanaka, Y., 2010. Molecular motors in neurons: transport mechanisms and roles in brain function, development, and disease. {\it Neuron}, 68(4), 610-638.

\bibitem{hirokawa2015} Hirokawa, N., Tanaka, Y., 2015. Kinesin superfamily proteins, KIFs: Various functions and their relevance for important phenomena in life and diseases. {\it Exp. Cell Res.}, 334(1), 16-25.

\bibitem{courson2015} Courson, D. S., Cheney, R. E., 2015. Myosin-X and disease. {\it Exp. Cell Res.}, 334(1), 10-15.

\bibitem{heissler2017} Heissler, S. M., Chinthalapudi, K., Sellers, J. R., 2017. Kinetic signatures of myosin-5B, the motor involved in microvillus inclusion disease. {\it J. Biol. Chem.}, 292(44), 18372-18385.

\bibitem{vera2019} Vera, C. D., Johnson, C. A., Walklate, J., Adhikari, A., Svicevic, M., Mijailovich, S. M., Combs, A. C., Langer, S. J., Ruppel, K. M., Spudich, J. A., Geeves, M. A., Leinwand, L. A., 2019. Myosin motor domains carrying mutations implicated in early or late onset hypertrophic cardiomyopathy have similar properties. {\it J. Biol. Chem.}, 294(46), 17451-17462.

\bibitem{andorfer2019} Andorfer, R., Alper, J. D., 2019. From isolated structures to continuous networks: A categorization of cytoskeleton-based motile engineered biological microstructures. {\it Wiley Interdisciplinary Reviews: Nanomedicine and Nanobiotechnology}, 11(4), e1553.


\bibitem{saper2019} Saper, G., Hess, H., 2019. Synthetic Systems Powered by Biological Molecular Motors. {\it Chemical reviews}.

\bibitem{albertsbook2013} Alberts, B., Bray, D., Hopkin, K., Johnson, A. D., Lewis, J., Raff, M., ... Walter, P., 2013. {\it Essential cell biology}. Garland Science.
      
\bibitem{luby1999} Luby-Phelps, K., 1999. Cytoarchitecture and physical properties of cytoplasm: volume, viscosity, diffusion, intracellular surface area. In {\it International review of cytology}, Vol. 192, pp. 189-221. Academic Press.
         
\bibitem{mullins1999} Mullins, R. D., Pollard, T. D., 1999. Structure and function of the Arp2/3 complex. {\it Curr. Opin. Struct. Biol.}, 9(2), 244-249.
 
\bibitem{goshima2008} Goshima, G., Mayer, M., Zhang, N., Stuurman, N., Vale, R. D., 2008. Augmin: a protein complex required for centrosome-independent microtubule generation within the spindle. {\it J. Cell Biol.}, 181(3), 421-429.

\bibitem{petry2013} Petry, S., Groen, A. C., Ishihara, K., Mitchison, T. J., Vale, R. D., 2013. Branching microtubule nucleation in Xenopus egg extracts mediated by augmin and TPX2. {\it Cell}, 152(4), 768-777.

\bibitem{mckenney2014} McKenney, R. J., Huynh, W., Tanenbaum, M. E., Bhabha, G., Vale, R. D., 2014. Activation of cytoplasmic dynein motility by dynactin-cargo adapter complexes. {\it Science}, 345(6194), 337-341.

\bibitem{spiliotis2008} Spiliotis, E. T., Hunt, S. J., Hu, Q., Kinoshita, M., Nelson, W. J., 2008. Epithelial polarity requires septin coupling of vesicle transport to polyglutamylated microtubules. {\it J. Cell Biol.}, 180(2), 295-303.

\bibitem{janke2011} Post-translational regulation of the microtubule cytoskeleton: mechanisms and functions. {\it Nat. Rev. Mol. Cell Biol.}, 12(12), 773-786.

\bibitem{cai2009} Cai, D., McEwen, D. P., Martens, J. R., Meyhofer, E., Verhey, K. J., 2009. Single molecule imaging reveals differences in microtubule track selection between Kinesin motors. {\it PLOS Biol.}, 7(10).

\bibitem{atherton2013} Atherton, J., Houdusse, A., Moores, C., 2013. MAPping out distribution routes for kinesin couriers. {\it Biol. Cell}, 105(10), 465-487.

\bibitem{schneider2015} Schneider, R., Korten, T., Walter, W. J., Diez, S., 2015. Kinesin-1 motors can circumvent permanent roadblocks by side-shifting to neighboring protofilaments. {\it Biophys. J}, 108(9), 2249-2257.

\bibitem{ali2007} Ali, M. Y., Krementsova, E. B., Kennedy, G. G., Mahaffy, R., Pollard, T. D., Trybus, K. M., Warshaw, D. M., 2007. Myosin Va maneuvers through actin intersections and diffuses along microtubules. {\it Proc. Natl. Acad. Sci. U.S.A.}, 104(11), 4332-4336.

\bibitem{ross2008} Ross, J. L., Shuman, H., Holzbaur, E. L., Goldman, Y. E., 2008. Kinesin and dynein-dynactin at intersecting microtubules: motor density affects dynein function. {\it Biophys. J}, 94(8), 3115-3125.

\bibitem{balint2013} B{\'a}lint, {\v{S}}.,   Vilanova, I.V.,   {\'A}lvarez, {\'A}.S. and Lakadamyali, M., 2017. 3D motion of vesicles along microtubules helps them to circumvent obstacles in cells. {\it J. Cell Sci.}, 130(11), 1904-1916.

\bibitem{miedema2017} Miedema, D. M., Kushwaha, V. S., Denisov, D. V., Acar, S., Nienhuis, B., Peterman, E. J., Schall, P., 2017. Correlation imaging reveals specific crowding dynamics of kinesin motor proteins. {\it Phys. Rev. X}, 7(4), 041037.

\bibitem{erickson2011} Erickson, R. P., Jia, Z., Gross, S. P., Yu, C. C., 2011. How molecular motors are arranged on a cargo is important for vesicular transport. {\it PLOS Comput. Biol}, 7(5).

\bibitem{verdeny2017} Verdeny-Vilanova, I.,   Wehnekamp, F.,   Mohan, N.,   {\'A}lvarez, {\'A}.S.,   Borbely, J.S.,   Otterstrom, J.J.,   Lamb, D.C. and Lakadamyali, M., 3D motion of vesicles along microtubules helps them to circumvent obstacles in cells. {\it J. Cell Sci.}, 130(11), 1904-1916.

\bibitem{bergman2018} Bergman, J. P., Bovyn, M. J., Doval, F. F., Sharma, A., Gudheti, M. V., Gross, S. P., ... , Vershinin, M. D., 2018. Cargo navigation across 3D microtubule intersections. {\it Proc. Natl. Acad. Sci. U.S.A.}, 115(3), 537-542.

\bibitem{gao2018} Gao, Y., Anthony, S. M., Yu, Y., Yi, Y., Yu, Y., 2018. Cargos rotate at microtubule intersections during intracellular trafficking. {\it Biophys. J}, 114(12), 2900-2909.

\bibitem{feng2018} Feng, Q., Mickolajczyk, K. J., Chen, G. Y., Hancock, W. O., 2018. Motor reattachment kinetics play a dominant role in multimotor-driven cargo transport. {\it Biophys. J}, 114(2), 400-409.

\bibitem{deeb2019}
  Deeb, S. K., Guzik-Lendrum, S., Jeffrey, J. D., Gilbert, S. P., 2019. The ability of the kinesin-2 heterodimer KIF3AC to navigate microtubule networks is provided by the KIF3A motor domain. {\it J. Biol. Chem.}, 294(52), 20070-20083.

\bibitem{hancock18} Mickolajczyk, K. J., Hancock, W. O., 2018. High-Resolution Single-Molecule Kinesin Assays at kHz Frame Rates. In {\it Molecular Motors}, pp. 123-138. Humana Press, New York, NY.

\bibitem{andrecka2018} Andrecka, J., Arroyo, J. O., Takagi, Y., de Wit, G., Fineberg, A., MacKinnon, L., ... , Kukura, P., 2015. Structural dynamics of myosin 5 during processive motion revealed by interferometric scattering microscopy. {\it Elife}, 4, e05413.

\bibitem{parmeggiani2003} Parmeggiani, A., Franosch, T., Frey, E., 2003. Phase coexistence in driven one-dimensional transport. {\it Phys. Rev. Lett.}, 90(8), 086601.

\bibitem{parmeggiani2004} Parmeggiani, A., Franosch, T., Frey, E., 2004. Totally asymmetric simple exclusion process with Langmuir kinetics. {\it Phys. Rev. E}, 70(4), 046101.

\bibitem{leduc2012} Leduc, C., Padberg-Gehle, K., Varga, V., Helbing, D., Diez, S., Howard, J., 2012. Molecular crowding creates traffic jams of kinesin motors on microtubules. {\it Proc. Natl. Acad. Sci. U.S.A.}, 109(16), 6100-6105.

\bibitem{embley2008}  Embley, B., Parmeggiani, A., Kern, N., 2008. HEX-TASEP: dynamics of pinned domains for TASEP transport on a periodic lattice of hexagonal topology. {\it J. Phys. Condens}, 20(29), 295213.

\bibitem{embley2009} Embley, B., Parmeggiani, A., Kern, N., 2009. Understanding totally asymmetric simple-exclusion-process transport on networks: Generic analysis via effective rates and explicit vertices. {\it Phys. Rev. E}, 80(4), 041128.

\bibitem{neri2011} Neri, I., Kern, N., Parmeggiani, A., 2011. Totally asymmetric simple exclusion process on networks. {\it Phys. Rev. Lett.}, 107(6), 068702.

\bibitem{neri2013} Neri, I., Kern, N., Parmeggiani, A., 2013. Modeling cytoskeletal traffic: an interplay between passive diffusion and active transport. {\it Phys. Rev. Lett.}, 110(9), 098102.

\bibitem{neri2013b} Neri, I., Kern, N., Parmeggiani, A., 2013. Exclusion processes on networks as models for cytoskeletal transport. {\it New J. Phys.}, 15(8), 085005.

\bibitem{raguin2013} Raguin, A., Parmeggiani, A., Kern, N., 2013. Role of network junctions for the totally asymmetric simple exclusion process. {\it Phys. Rev. E}, 88(4), 042104.

\bibitem{parmeggiani2014} Parmeggiani, A., Neri, I., Kern, N., 2014. Modelling Collective Cytoskeletal Transport and Intracellular Traffic. In {\it The Impact of Applications on Mathematics}, pp. 1-25. Springer, Tokyo.

\bibitem{schadschneider2000} Chowdhury, D., Santen, L., Schadschneider, A., 2000. Statistical physics of vehicular traffic and some related systems. {\it Phys. Rep.}, 329(4-6), 199-329.

\bibitem{derrida1993} Derrida, B., Evans, M. R., Hakim, V., Pasquier, V., 1993. Exact solution of a 1D asymmetric exclusion model using a matrix formulation. {\it J. Phys. A: Mathematical and General}, 26(7), 1493.

\bibitem{schutz1993} Sch{\"u}tz, G. and Domany, E., 1993. Phase transitions in an exactly soluble one-dimensional exclusion process. {\it J. Stat. Phys.}, 72(1-2), 277-296.

\bibitem{chou2011} Chou, T., Mallick, K., Zia, R. K. P., 2011. Non-equilibrium statistical mechanics: from a paradigmatic model to biological transport. {\it Rep. Prog. Phys.}, 74(11), 116601.

\bibitem{kolomeisky2005} Pronina, E., Kolomeisky, A. B., 2005. Theoretical investigation of totally asymmetric exclusion processes on lattices with junctions. {\it J. Stat. Mech.: Theory Exp}, 2005(07), P07010.

\bibitem{muller2008} M{\"u}ller, M.J.I., Klumpp, S. and Lipowsky, R., 2008. Tug-of-war as a cooperative mechanism for bidirectional cargo transport by molecular motors., {\it Proc. Natl. Acad. Sci. U.S.A.}, 105(12), 4609-4614.

\bibitem{reichenbach2007} Reichenbach, T., Frey, E., Franosch, T., 2007. Traffic jams induced by rare switching events in two-lane transport. {\it New J. Phys.}, 9(6), 159.

\bibitem{melbinger2011} Melbinger, A., Reichenbach, T., Franosch, T., Frey, E., 2011. Driven transport on parallel lanes with particle exclusion and obstruction. {\it Phys. Rev. E}, 83(3), 031923.

\bibitem{howard2003} Howard, J., Hyman, A. A., 2003. Dynamics and mechanics of the microtubule plus end. {\it Nature}, 422(6933), 753-758.

\bibitem{conde2009} Conde, C. and C{\'a}ceres, A., 2009. Microtubule assembly, organization and dynamics in axons and dendrites. {\it Nat. Rev. Neurosci.}, 10(5), 319-332.

\bibitem{okada2015} Okada, N., Sato, M., 2015. Spatiotemporal regulation of nuclear transport machinery and microtubule organization. {\it Cells}, 4(3), 406-426.

\bibitem{goodson2018} Goodson, H. V., Jonasson, E. M., 2018. Microtubules and microtubule-associated proteins. {\it CSH Perspect. Biol.}, 10(6), a022608.


\bibitem{Lombardo2017}
Lombardo, A. T., Nelson, S. R., Ali, M. Y., Kennedy, G. G., Trybus, K. M., Walcott, S., Warshaw, D. M., 2017. Myosin Va molecular motors manoeuvre liposome cargo through suspended actin filament intersections in vitro. {\it Nat. Commun.}, 8(1), 1-9.

\end{thebibliography}
\end{document}